\documentclass[prb,showpacs,floatfix,superscriptaddress,amsmath,amssymb,twocolumn]{revtex4-1}
\usepackage[english]{babel} 
\usepackage{bm}
\usepackage{graphicx}
\usepackage{amsmath}
\usepackage{amssymb}
\usepackage{wasysym}
\usepackage{comment}
\usepackage{xcolor}
\usepackage{tikz}
\usepackage{adjustbox}
\usepackage{booktabs}
\usepackage{array}

\makeatletter
\newcommand*{\encircled}[1]{\relax\ifmmode\mathpalette\@encircled@math{#1}\else\@encircled{#1}\fi}
\newcommand*{\@encircled@math}[2]{\@encircled{$\m@th#1#2$}}
\newcommand*{\@encircled}[1]{%
	\tikz[baseline,anchor=base]{\node[draw,circle,outer sep=0pt,inner sep=.2ex] {#1};}}
\makeatother

\usepackage[hidelinks]{hyperref}

\begin{document}

\title{Combined approach to analyze and classify families of classical spin liquids}

        \author{N. Davier}
		\email[]{davier@irsamc.ups-tlse.fr}
		\affiliation{Laboratoire de Physique Th\'eorique, Universit\'e de Toulouse, CNRS, UPS, France}
		 
            \author{F. A.  G\'omez Albarrac\'in}
		\email[]{albarrac@fisica.unlp.edu.ar}
		\affiliation{Instituto de F\'isica de L\'iquidos y Sistemas Biol\'ogicos, CONICET, Facultad de Ciencias Exactas, Universidad Nacional de La Plata, 1900 La Plata, Argentina}
\affiliation{Departamento de Ciencias B\'asicas, Facultad de Ingenier\'ia, UNLP, La Plata, Argentina}

            \author{H. D.  Rosales}
		\email[]{rosales@fisica.unlp.edu.ar}
		\affiliation{Instituto de F\'isica de L\'iquidos y Sistemas Biol\'ogicos, CONICET, Facultad de Ciencias Exactas, Universidad Nacional de La Plata, 1900 La Plata, Argentina}
\affiliation{Departamento de Ciencias B\'asicas, Facultad de Ingenier\'ia, UNLP, La Plata, Argentina}
		
		\author{P. Pujol}
		\email[]{pierre.pujol@irsamc.ups-tlse.fr}
		\affiliation{Laboratoire de Physique Th\'eorique, Universit\'e de Toulouse, CNRS, UPS, France}
		
		\date{\today}
		
\begin{abstract} 
        Classical spin liquids have been a very active subject of research in the last few years. A very rich variety of cases have been shown to exist, including short-range and algebraic spin liquids displaying dipolar correlations at zero temperature. In this article, we investigate different families of classical spins liquids by combining analytical techniques and Monte Carlo simulations. Our study relies on the Luttinger-Tisza approximation (LTA), which is associated with the constraint vector function in momentum space, whose general properties allow for a classification of different spin liquids. We show that the general properties of the LTA provide a framework for identifying and accurately characterizing the different types of spin liquids in different geometries. We apply our approach to three different families of spin liquids defined on the checkerboard and kagome lattices, which exhibit a remarkable range of situations, including various cases of algebraic and short-range spin liquids. Additionally, we analyze the effective Gauss law emerging from different kinds of spin liquids and identify states that exhibit additional degeneracy lines. The presence of spin-liquid phases and pinch-point singularities are confirmed by Monte Carlo simulations validating our approach. Our study opens up avenues of research in the study of spin liquids, exploring algebraic spin liquids with higher-rank gauge fields and as critical points dividing different types of classical spin liquids.

\end{abstract}
		
		\maketitle

\section{Introduction}

Classical frustrated antiferromagnets are known to show exotic low-temperature behavior, for example, the absence of order and an extensive zero temperature entropy, the simplest example has been provided by the plain Ising antiferromagnet in the triangular lattice \cite{Wannier-Ising}. The Heisenberg antiferromagnet in the checkerboard and the kagome lattices are other examples of a system with extensive zero-temperature entropy. Of particular interest is the behavior of, for example, correlation functions at zero temperature, as it is solely governed by the entropic properties of the lowest energy set of configurations. The nature of this lowest energy set of configurations, sometimes also called ground-state manifold, may present interesting properties implying, for example, algebraically decaying correlations. A situation where this occurs is when all the lowest energy  configurations satisfy a conservation property that can be assimilated to a charge-free Gauss Law. In those cases, because of entropic reasons, the probability weight of coarse-grained configurations is built from an effective free energy with a Maxwell form \cite{Gauss_pyro, Henley2005}. For spin systems, it was first realized in the case of the three-dimensional spin-ice system in the pyrochlore lattice \cite{Anderson_pyro, Pol_ice} later studied for Heisenberg spins \cite{kagome_largeN, Gauss_pyro, Henley2005}. This also applies to the two-dimensional counterparts which are the checkerboard  and  the kagome lattices. The algebraic correlations arise in this case from the propagator of a ``photon'' which is not screened by charge proliferation as long as only the lowest energy set of configurations is taken into account. In general, this zero temperature behavior is also captured at non-zero temperature, with for example MC analysis, provided that there is no Order By Disorder (OBD) mechanism \cite{OBDVillain} that selects a subspace within the ground-state manifold.   
The existence of algebraic correlations at zero temperature is not nevertheless a generic feature of those highly degenerate systems as there are also known examples having only exponentially decaying correlations \cite{Rehn_Moessner_2017}, and so excluding the possibility of an unscreened photon. 

In this article, we investigate the spin liquid behavior of three families of classical Heisenberg systems on two-dimensional lattices - specifically, the checkerboard and kagome lattices. These families are defined by including longer-range couplings beyond nearest neighbors, with the values of these couplings spanning the parameter space of the models. To maintain the spin-liquid nature of the systems, we write the Hamiltonian as a sum of the squares of the magnetizations of effective plaquettes. By exploring the phase diagrams of the models as we vary the parameters, we uncover a rich variety of behaviors, including algebraic spin liquids that are associated with a Gauss Law satisfied by vector or tensor gauge fields. We also find short-range spin liquid phases that are separated by critical points corresponding to algebraic spin liquids. Importantly, we confirm all of our results through Monte Carlo simulations.  

In sec. \ref{sec: Analytical and numerical methods} we describe the kind of Hamiltonians we consider and the analytical and numerical techniques used all along the article. In sec. \ref{sec: Checkerboard} we treat as a first example the case of a generalized checkerboard lattice. This case turns out to have an algebraic behavior with an associated Gauss Law in all the regions of its phase diagram. The second example is given in sec. \ref{sec: Kagome} and it is built from the kagome lattice. This model  also shows an algebraic behavior with an associated Gauss law for every value of the parameters although it presents a different behavior with temperature: in the checkerboard case, the specific heat normalized to the expected value for a non-liquid system is always less than 1  (indicating the presence of zero and soft modes), while in the second example, OBD is at play at lower temperatures. The sec. \ref{sec: Corner sharing hexagons} is devoted to the last example, which is built as  corner sharing hexagons lattice. In contrast to the two previous models, this system has short-range spin liquid phases separated by critical points with algebraic behavior. At this stage, it is worth mentioning that one key ingredient to analyze the low-temperature behavior of those systems is what we call the constraint vector, which we discuss below, and whose nature is different for the three families of systems that we analyze here.   

\section{Approach and Methods}
\label{sec: Analytical and numerical methods}

The aim of this study is to investigate the various types of classical spin liquids and their general properties that arise from the Heisenberg model defined in generalized lattices. Three types of lattices are considered: the checkerboard lattice, the kagome lattice represented as corner-sharing triangular plaquettes, and the kagome lattice represented as corner-sharing hexagonal plaquettes. The antiferromagnetic Heisenberg models defined in these lattices are known to be highly frustrated systems giving rise to a very rich phenomenology. Here, we define a generalization for each of these cases by adding longer-range couplings between the spins but preserving the fundamental property of the Hamiltonian to be expressed as the sum of the square of the magnetization of plaquettes. The primary differentiating factor between these models is the nature of the constraint vector function, which is described in detail in subsection B and was introduced in Ref. [\onlinecite{Henley2005, Benton_Moessner_2021}]. 

In this study, we focus on systems described by a Hamiltonian that can be expressed as a sum of clusters of magnetic sites on a lattice. Specifically, we consider a lattice consisting of classical spins $\mathbf{S}_i$, which correspond to three-component unit vectors defined on each lattice site $i$. The Hamiltonian takes the form
\begin{equation}
    H = \frac{J}{2} \sum_{p} \bm{\mathcal{S}}_p ^2 \label{eq : General Hamiltonian}
\end{equation}
where the sum is made over clusters, or plaquettes labeled $p$ and 
\begin{equation}
    \bm{\mathcal{S}}_p = \sum_{i \in p} \eta_i\mathbf{S}_i, \label{eq : plaquette}
\end{equation}
with $\eta_i$ being real coefficients. These coefficients allow to tune continuously the Hamiltonian and thus the interactions, but conserving the general Hamiltonian structure of Eq.~(\ref{eq : General Hamiltonian}). By construction, what we call the effective spin of the plaquette $p$, $\bm{\mathcal{S}}_p$, must be zero for all the plaquettes in order to minimize the energy. Depending on the chosen plaquettes, this Hamiltonian implies exchange interactions at different order of neighbors with specific ratios, related to the number of plaquettes where the bond is present.

To comprehensively analyze these models and investigate the diverse types of spin liquids, we employ for each of them both analytical and numerical approaches, which involve  various complementary steps:
\begin{itemize}
\item We first study the band structure of the models using the \textbf{Luttinger-Tisza approximation (LTA)}, placing special emphasis on the points where dispersive bands touch flat bands. By applying LTA, we can predict potential spin liquid regimes and gain a deeper understanding of the system's behavior.

\item We study the properties of the \textbf{Constraint Vector}, whose characteristics are different in each of the chosen models, and argue how its analysis gives us important information about the properties of the system at very low temperatures, as for example the shape of the structure factors. Moreover, as was pointed out in Ref. [\onlinecite{Benton_Moessner_2021}], the study of its topological properties can also give us information about the nature of the spin liquids and the transitions that may occur between different spin liquid phases. Also, following Henley\cite{Henley2005}, we perform a \textbf{Projective Analysis} by connecting the spin correlation functions to a projector in the space orthogonal to \textbf{Constraint Vector}.

\item Then, we resort to \textbf{Monte Carlo simulations (MC)}, to study the effect of temperature and calculate key parameters to unveil the different types of spin liquids.

\item Finally, we show that in many (although not all) cases, the lowest energy configurations manifold of the system are characterized by a \textbf{Gauss' Laws}, which is associated to the presence of pinch points in the structure factors \cite{Henley2005}. We also observe for some critical cases other emergent phenomena, such as \textbf{higher rank pinch points}. 
\end{itemize}
Before delving into the details of the three models we have analyzed, it is worthwhile to provide some general context regarding the analytical and numerical approaches we employed.

\subsection{Luttinger-Tisza Approximation}

The Luttinger-Tisza approximation\cite{LT1,LT2,kaplan_2007} (LTA), also known as the spherical approximation, is a powerful tool to provide an initial characterization of the classical ground states of quadratic Hamiltonians. This approach begins by defining the Fourier transforms of the spins $\mathbf{S}^m_{\mathbf q}$ as

\begin{equation}
 \mathbf{S}^m_\mathbf{q} =\frac{1}{N_c} \sum_{i} \mathbf{S}^m_i e^{-i\mathbf{q}\cdot\mathbf{r}_{i,m}}
 \label{Eq:Sq_Fourier}
\end{equation}
where we have assumed that the lattice has $N_c$ unit cells, the index $m$ index the sublattice and the index $i$ run over all unit cells. The vector $\mathbf{r}_{i,m}$ gives the position of the site from sublattice $m$ among the unit cell $i$. In order to get the $T=0$ spin configurations, within the LTA scheme, instead of imposing the local constraint in spin length $\|\mathbf{S}_i\|=1$ is replaced by the softer global constraints $\sum_i\|\mathbf{S}_i^m\|^2=N_c\,S^2$, one for each sublattice. Within this less restrictive approximation, if the system is translation invariant, we can diagonalize the Hamiltonian by taking the Fourier transform of the spins as Eq.~(\ref{Eq:Sq_Fourier}). So, for a general classical Heisenberg model of magnetic moments coupled by exchange interactions $J_{ij;mn}$, $H =\sum_{ij} \sum_{mn}J_{ij;mn}\mathbf{S}^{m}_i\cdot\mathbf{S}^{n}_j$, it is possible to rewrite the Hamiltonian as 
\begin{equation}
    H =\frac{1}{N_c}\sum_{m,n}\sum_{\mathbf{q}}M_{mn}(\mathbf{q})\, \mathbf{S}^m_{\mathbf{q}}\cdot\mathbf{S}^n_{-\mathbf{q}} \label{eq: H general momentum space}
 \end{equation}
where $M_{mn}(\mathbf{q})=[\mathbf{M}(\mathbf{q})]_{mn}$ corresponds to the Fourier transformation of the exchange interactions
\begin{equation}
    M_{mn}(\mathbf{q}) =\sum_{\mathbf{r}_{j}-\mathbf{r}_i} J_{mn}(\mathbf{r}_{j}-\mathbf{r}_i)\, e^{-i\mathbf{q}\cdot(\mathbf{r}_{j,n}-\mathbf{r}_{i,n})}
 \end{equation}
with $\mathbf{r}_{j}-\mathbf{r}_i$ vectors representing lattice translations linking different unit cells, and where the $\sum_{\mathbf{q}}$ runs over all wave vectors in the first Brillouin zone. The eigenvalues $\{\varepsilon_{m}(\mathbf{q})\}$ of the matrix $\mathbf{M}(\mathbf{q})$ correspond to the energy bands of the model while the ground-state configuration (at $T=0$) is associated with the minima of the lowest band $\varepsilon_{m}(\mathbf{q})$  which defines the ordering wave-vectors $\mathbf{q^*}$. For Bravais lattices, the ground states of the Hamiltonian can always be constructed as a linear combination of the eigenvalues obtained from the LTA. For non-Bravais lattices, such as the ones considered in this work, the LTA provides a low-energy boundary for the ground states. It is also an indicator of frustration and ground-state degeneracy. Thorough this manuscript we will exploit this tool in connection with the constraint vector, which we describe in the next subsection.

\subsection{The Constraint Vector}

For a system with translational invariance, and from the parameters of the generalized plaquette defined in Eq.~(\ref{eq : plaquette}), one can define what we call the constraint vector function \cite{Henley2005, Benton_Moessner_2021} in momentum space:
\begin{equation}
        L^m_\mathbf{q} = \sum_{i, m \in p} \eta_i\,e^{i \mathbf{q}\cdot( \mathbf{r_c} - \mathbf{r}_{i,m} )}.
\end{equation}
where the sum is made over sites of sublattice $m$ belonging to a generalized plaquette $p$. $\mathbf{q}$ is the momentum and $\mathbf{r}_c$ is the real-space vector position of the plaquette center\cite{Benton_Moessner_2021}. For the special case of a Hamiltonian with the structure in Eq.~(\ref{eq : General Hamiltonian}), this definition allows us to rewrite the Hamiltonian as 
\begin{equation}
    H = \frac{J}{2} \sum_{m,n}\sum_{\mathbf{q}}(L^m_\mathbf{q} L^n_{-\mathbf{q}})\, \mathbf{S}^m_{\mathbf{q}}\cdot\mathbf{S}^n_{-\mathbf{q}} \label{eq: ref LT}
 \end{equation}
where $\mathbf{S}^n_{-\mathbf{q}}$ are the Fourier transform of the spins $\mathbf{S}_i$. 
This notation is reminiscent of the above-mentioned LTA of a generic spin Hamiltonian in momentum space (see Eq.~(\ref{eq: H general momentum space})). It appears that in the specific case of cluster Hamiltonian (\ref{eq : General Hamiltonian}) there will be an intimate connection between the constraint vector and the band structure. 

\begin{figure}[h]
    \centering
    \includegraphics[width = 0.7\linewidth]{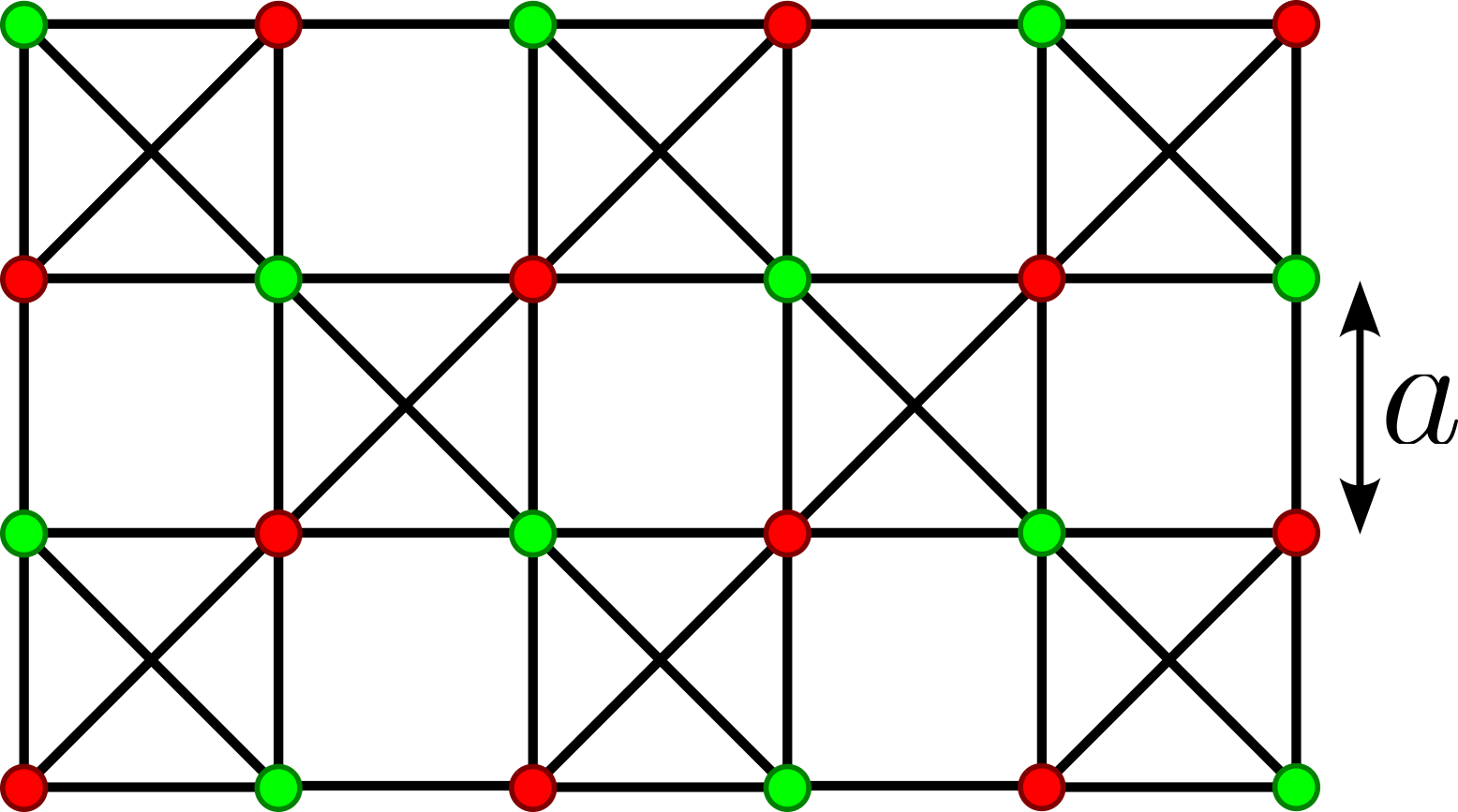}
    \caption{Checkerboard lattice. The lattice spacing is denoted by $a$. This lattice possesses two sublattices, depicted by red and green dots. }
    \label{fig: Ckb simple}
\end{figure}

Consider the example of the checkerboard lattice, depicted on Fig.~\ref{fig: Ckb simple}, with antiferromagnetically interacting Heisenberg spins placed on each vertex. For this system, the interaction matrix writes
\begin{equation}
    \mathbf{M}(\mathbf{q}) = J\begin{pmatrix}
        \cos \left[a(q_x+q_y)\right]  &  \cos (aq_x) + \cos (aq_y) \\
        \cos (aq_x) + \cos (aq_y) & \cos \left[a(q_x-q_y)\right]
    \end{pmatrix} 
\end{equation}
where the constraint vector is 
\begin{equation}
    \mathbf{L}(\mathbf{q}) = \begin{pmatrix}
        2 \cos\left[a(q_x+q_y)/2\right] \\
        2 \cos\left[a(q_x-q_y)/2\right]
    \end{pmatrix}.
\end{equation}
Thus it is easy to show that $M_{mn}(\mathbf{q}) = \frac{J}{2} L_m(\mathbf{q})L_n(\mathbf{q}) - 1\!\!1$ using the identity $2\cos(a+b)\cos(a-b) = \cos a + \cos b$.

Note that since $M_{mn}(\mathbf{q})$ is composed of a single vector $\mathbf{L}_\mathbf{q}$, the number of dispersive and flat bands can be easily predicted from the properties of $\mathbf{L}_\mathbf{q}$. If $\mathbf{L}_\mathbf{q}$ is a $n$-component real vector, one can choose a basis where all its components are zero except one. This indicates the presence of $n-1$ flat bands associated with the lowest energy topped by a unique dispersive band having a dispersion $\varepsilon(\mathbf{q}) = \frac{J}{2} \|\mathbf{L}_\mathbf{q}\|^2$. If $\mathbf{L}_\mathbf{q}$ is a $n$-component complex vector, then one expects $n-2$ flat bands.
The two cases reveal to be different and it is convenient to describe the two possibilities separately, showing in each case why the constraint vector is an efficient tool for the identification of pinch points. Moreover, from the analysis that we detail below, it becomes clear how the $T \to 0$ and non-zero temperature behavior \cite{Halfmoons} are related.

\subsection{Unique Constraint Vector and Projective Analysis}

Let us consider first the case where the cluster Hamiltonian is defined on a unique type of plaquette. For each of these clusters $p$, the condition $\bm{\mathcal{S}}_p=0$ translates in reciprocal space into the condition 
\begin{equation}
    \mathcal{S}^\alpha(\mathbf{q}) = \sum_m L^m_{\mathbf{q}} S^{m,\alpha}_{\mathbf{q}} =0
    \label{eq: ground state constraint}
\end{equation}
with $m$ indexing sublattices and $\alpha=x,y,z$ referring to the spin component.
In this view, and following Henley's arguments \cite{Henley2005}, it appears that for $T \to 0$, the correlation functions $\langle \mathbf{S}^{m}_{\mathbf{-q}}\cdot\mathbf{S}^{n}_{\mathbf{q}} \rangle$ must be proportional to the projector in the space orthogonal to $\mathbf{L}_{\mathbf{q}}$, defined as
\begin{equation}
    \mathbf{\Pi} = 1 - \mathbf{L} \frac{1}{\|\mathbf{L}\|^2} \mathbf{L}^\dagger,
    \label{eq: projector unique real Lq}
\end{equation}
and is manifestly singular when $\|\mathbf{L}\|^2$ vanishes.  Within this Projective Analysis of correlation functions, as we argue below, this situation typically corresponds to the presence of a pinch point and the corresponding algebraic correlations in real space. This directly explains why, for centrosymmetric systems corresponding to a real-valued constraint vector, the topological analysis of the constraint vector appears to be useful\cite{Benton_Moessner_2021}. It indeed allows identifying the situations for which the vector constraint vanishes.
In the context of the LTA, this implies that pinch points will be observed each time the dispersive band touches the flat band as the dispersion law is proportional to the square of the constraint vector norm $\varepsilon(\mathbf{q}) \equiv \frac{J}{2}\|\mathbf{L}_{\mathbf{q}}\|^2$. 

The structure of the contact point determines the type of pinch point we observe. Let us first see the most typical situation in which the dispersion around the contact point is quadratic
\begin{equation}
    \varepsilon(\mathbf{q}) \propto \|\mathbf{q}\|^2 .
\end{equation}
This case corresponds to the usual pinch points, associated with the correlation functions 
\begin{equation}
    \langle \mathbf{S}^{m}_{\mathbf{-q}}\cdot\mathbf{S}^{n}_{\mathbf{q}}\rangle \sim \delta_{mn} - \frac{q_m q_n}{\|\mathbf{q}\|^2}
\end{equation}
where we have assumed a linear behavior for two of the components of the vector constraint
\begin{equation}
    L^n_{\mathbf{q}} = a_iq_i
\end{equation}
since its norm is quadratic.

Let us now consider the case where the dispersion around a contact point reveals to be quartic and not quadratic. The dispersion relation is thus
\begin{equation}
    \varepsilon(\mathbf{q}) \sim \|\mathbf{L}_{\mathbf{q}}\|^2 =  \alpha \|\mathbf{q}\|^4 + 2\beta q_x^2q_y^2
\end{equation}
indicating a behavior for the correlation functions like 
\begin{equation}
    \langle \mathbf{S}^{m}_{-\mathbf{q}}, \mathbf{S}^{n}_{\mathbf{q}} \rangle \propto \delta_{mn}  - \frac{(q_m)^2 (q_n)^2}{\alpha \|\mathbf{q}\|^4 + 2\beta q_x^2q_y^2}
\end{equation}
which corresponds to a higher rank four-fold symmetric pinch point\cite{Prem_2018}. 

\begin{figure}
    \centering
    \includegraphics[width =0.8 \columnwidth]{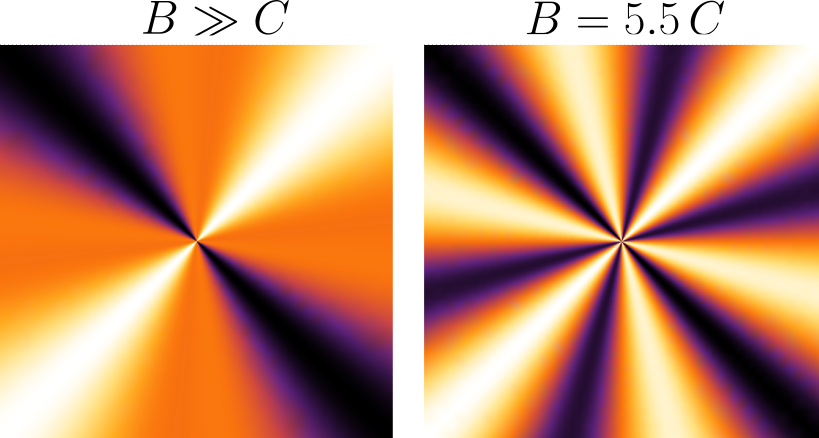}
    \caption{Different types of pinch points structures associated with rank three divergentless tensors. The case $B \gg C$ looks like usual pinch points but flattened along the direction perpendicular to the two arms. For $B \gtrsim C$ the pinch points possess 6 arms, which contrast depends on the ratio $B/C$. The case $C \gg B$ is not illustrated as it simply corresponds to usual pinch points.}
    \label{fig: Rank 3 pinch points structures}
\end{figure}

If the local dispersion reveals to be sextic
\begin{equation}
    \varepsilon(\mathbf{q}) \simeq \alpha^2 ( q_x^6 + q_y^6) + 2 \beta (q_x^4q_y^2 + q_x^2q_y^4),
\end{equation}
the situation is more complex. In this case, the structure factor can be generally described by a function of the form
\begin{equation}
    S(\mathbf{q}) \sim A + B \left(\frac{q^xq^y}{q_x^2+q_y^2}\right)^3 -C \frac{q^xq^y}{q_x^2+q_y^2}. 
    \label{eq: structure pp rang 3}
\end{equation}
where $A$, $B$ and $C$ are three real constants. The ratio $B/C$ determines the aspect of the pinch point observed, see Fig.~ \ref{fig: Rank 3 pinch points structures}. For $C \gg B$ pinch points look like usual pinch points since the structure factor becomes similar to the one associated with a quadratic dispersion. However, for $B \gtrsim C$,  pinch points present a six-leg structure, easily recognizable. For $B \gg C$ the pinch points look again like regular pinch points but flattened. 

\subsection{Complex Conjugate Constraint Vectors and Projective Analysis}
\label{subsec: complex conjugate constraint vectors}
 We focus now on systems where there exist two ground state conditions as in Eq.~(\ref{eq: ground state constraint}), involving two different constraint vectors which are simply complex conjugate. These conditions correspond to the case of lattices such as the pyrochlore lattice\cite{Henley2005} or the kagome lattice with first neighbors couplings, or its generalization that we mention here in the section of corner-sharing triangles. In this kind of situation, the system has different types of plaquettes and equation (\ref{eq : General Hamiltonian}) must be modified to write the sum over the different kinds of plaquettes. In the simplest case, the elementary clusters are triangles, which do not possess a central point symmetry. Because there are up and down triangles, there are two relations like Eq.~(\ref{eq: ground state constraint}), one for each type of triangle, corresponding by symmetry to complex conjugate constraint vectors. This situation is similar to the tetrahedra conditions in the pyrochlore lattice\cite{Henley2005}. 

For these systems, the lowest energy constraint doubles into $\mathbf{L}\cdot \mathbf{S}^\alpha =0$ and $\mathbf{L}^*\cdot \mathbf{S}^\alpha =0$, one for each vector constraint (and for each spin component $\alpha$).  Note that the components of the present vectors refer to sublattices and not spin components. This means that, for the Projective Analysis,  the correlation functions $\langle \mathbf{S}^{m}_{\mathbf{-q}}\cdot\mathbf{S}^{n}_{\mathbf{q}} \rangle$ are now proportional to the projector into the space orthogonal to both $\mathbf{L}_{\mathbf{q}}$ and $\mathbf{L}^*_{\mathbf{q}}$. To build this projector, we define a matrix $\mathbf{M}$ made of $\mathbf{L}$ and $\mathbf{L}^*$ as columns and define the projector into the subspace orthogonal to $\mathbf{M}$ as 
\begin{equation}
    \mathbf{\Pi} \equiv 1 - \mathbf{M} (\mathbf{M}^\dagger \mathbf{M} )^{-1}\mathbf{M}^\dagger,
\end{equation}
where 
\begin{equation}
    \mathbf{M}^\dagger \mathbf{M} = 
    \begin{pmatrix}
        \|\mathbf{L}\|^2 &  Q^* \\
        Q  & \|\mathbf{L}\|^2 
    \end{pmatrix}
\end{equation}  
and
\begin{equation}
    Q(\mathbf{q}) = Q^* (\mathbf{-q}) =  \mathbf{L} \cdot \mathbf{L} = \sum_{m} \big(L^m_{\mathbf{q}}\big)^2 
\end{equation}
is a complex scalar function of momentum $\mathbf{q}$. This leads to 
\begin{equation}
    (\mathbf{M}^\dagger \mathbf{M})^{-1} = \frac{1}{\|\mathbf{L}\|^4 - |Q|^2} \begin{pmatrix}
        \|\mathbf{L}\|^2 &  -Q^* \\
        -Q  & \|\mathbf{L}\|^2
    \end{pmatrix},
\end{equation}
implying that there are singularities anytime that we have 
\begin{equation}
    \|\mathbf{L}\|^4 - |Q|^2 = 0.
\end{equation}
Let us now consider the LTA, where the Hamiltonian in reciprocal space reads
\begin{equation}
    H = \sum_{\mathbf{q}} \left[ L^n_\mathbf{q} L^m_\mathbf{-q} + (L^n_\mathbf{q} L^m_\mathbf{-q})^* \right] \mathbf{S}^m_\mathbf{q}\cdot \mathbf{S}^n_\mathbf{q}
\end{equation}
with the first $\mathbf{L}_\mathbf{q} $ product accounting for up clusters and the conjugate one for down clusters. 
Again, one can always choose a basis in which the two first axis generate the plane spanned by the real and complex part of $\mathbf{L}$. In this basis, the transformed $\mathbf{L}$, which we note as $\mathbf{l}$, has thus only two components. The non-zero part of the LTA matrix can then be written as 
\begin{equation}
    \mathbf{m} 
    = \begin{pmatrix}
        2 |l_1|^2 & l_1l_2^* + l_1^*l_2 \\
        l_1l_2^* + l_1^*l_2  & 2 |l_2|^2 
    \end{pmatrix} ,
\end{equation}
indicating that the energies associated with the dispersive bands are simply
\begin{equation}
    \varepsilon_\pm = \|\mathbf{L}\|^2 \pm |Q|
\end{equation}
where we used the invariance of the scalar products under the rotation $\mathbf{L} \to \mathbf{l}$. We thus see that, once again, the observation of pinch points is related to the existence of a contact point between one of the dispersive bands and the flat ones.\\

At this point, it is noteworthy to make a remark regarding the analysis of the topological properties of the constraint vector $\mathbf{L}$. As we saw above, for a singularity to appear, the quantity that must be zero is not the norm of the constraint vector itself, but $(\mathbf{L}\cdot \mathbf{L}^*)^2 - |\mathbf{L}\cdot \mathbf{L}|^2$. This quantity can be rewritten, splitting real and imaginary parts of the constraint vector as 
\begin{equation}
    \|\mathbf{L}\|^4 - |Q|^2 = 4 \left[ \| R\{\mathbf{L}\}\|^2 \| I\{\mathbf{L}\}\|^2 -  \left(R\{\mathbf{L}\} \cdot I \{\mathbf{L}\} \right)^2 \right].
\end{equation}
This means that there is a singularity each time that either the real part or the imaginary part of the constraint vector vanishes, or they become proportional to each other. Therefore, the relevant vector field for the analysis is the real vector $\mathbf{L}_{\times}$ defined as:
\begin{equation}
    \mathbf{L}_{\times}(\mathbf{q}) = i~\mathbf{L_q}\times \mathbf{L^*_q}. 
\end{equation}
Here the square of the norm of $\mathbf{L}_{\times}$ is precisely equal to $\|\mathbf{L}\|^4 - |Q|^2$. For systems with $n$ distinct constraint vectors, a singularity in the structure factor is observed when the determinant of the matrix $\mathbf{M}^\dagger\cdot\mathbf{M}$ vanishes, where $\mathbf{M}$ is the matrix formed by arranging the $n$ constraint vectors as columns. This condition corresponds to the presence of linear dependence between any two of the constraint vectors\cite{Benton_Moessner_2021}. The topological analysis should thus be performed on the $\begin{pmatrix}n \\ 2\end{pmatrix}$ vectors fields defined as vector products of a pair of constraint vectors.

\subsection{Low-temperature emergent Gauss tensors}
\label{subsec: Low-temperature emergent Gauss tensors}

It is established in the literature\cite{Henley2005} that an effective low-energy Gauss law is connected with the presence of pinch points in the structure factor. This effective low-energy Gauss law may be described in terms of a gauge theory, {\it i.e.} the possibility to define a lattice vector field whose total flux at each vertex of the lattice is zero within the lowest energy configuration manifold. Although the converse is not obvious, it is nevertheless natural to ask if a singularity in the constraint vector as the ones we have discussed above implies the existence of a vector field with zero divergence in the limit as $T$ approaches zero. Under the assumption that no OBD effect takes place, the LTA can bring some light on this issue by considering the system at non-zero but very low temperature and at scales smaller than the thermal correlation length. \\

Consider a contact point $\mathbf{q}_0$ in the lowest dispersive band $\varepsilon(\mathbf{q})$ and the flat bands. These points play a crucial role in studying the large-scale behavior of correlation functions. By expanding the energy around the contact point $\mathbf{q} = \mathbf{q}_0$, one can generally obtain, a first-order approximation as

\begin{equation}
    \varepsilon(\mathbf{q}) \simeq \alpha^2 (q_x^2 +  q_y^2),
    \label{eq: quadratic pinch point}
\end{equation}
where $\mathbf{q}$ represents the deviation of momentum from the contact point $\mathbf{q}_0$, and $\alpha$ is a positive real coefficient and an axis rescaling may be necessary to make the expression isotropic. This case corresponds to the most common pinch point structure, which can be associated with a gauge field $\mathbf{E}$ that satisfies the relation
\begin{equation}
    \mathbf{q}\cdot \mathbf{E} = 0.
\end{equation}
This gauge field can be simply constructed as 
\begin{equation}
    E_x(\mathbf{q}) = -\alpha\, q_y\, \phi(\mathbf{q}), \hspace{0.7cm} E_y(\mathbf{q}) = \alpha\, q_x\,\phi(\mathbf{q}),
\end{equation}
where $\phi(\mathbf{q})$ is a coarse-grained version of the Fourier transform of a spin component. The associated energy functional describing the low energy physics is
\begin{equation}
    \begin{split}
        F &\simeq \int d^2r\, E_i E^i = 
        \int d^2q\, E_i(\mathbf{q}) E^i(-\mathbf{q}) \\ 
        &= \int d^2q\,\varepsilon(\mathbf{q}) |\phi(\mathbf{q})|^2.
    \end{split}
\end{equation}
This low-energy functional can be seen as a local development of the coarse-grained version of Hamiltonian in Eq.~(\ref{eq: H general momentum space}). However, understanding how this emergent Gauss law manifests at the lattice level with the spin degrees of freedom is a challenging and complex question that cannot be answered in a general way, but rather must be analyzed on a case-by-case basis for each specific system.  

Furthermore, at certain critical points or lines in the phase diagram, higher-rank pinch points can also appear as contact points. In these cases, the band dispersion can be described by a quartic development as
\begin{equation}
    \varepsilon(\mathbf{q}) \simeq \alpha^2 ( q_x^4 + q_y^4) + 2 \beta\,q_x^2q_y^2, 
    \label{eq: quartic pinch point}
\end{equation}
with $\alpha$ and $\beta$ being real positive coefficients. In this case, the underlying gauge structure involves a symmetric rank-2 tensor $E_{ij}$ satisfying $\partial_i E_{ij}=0$ \cite{Pretko_2017,Pretko_2_2017} which is translated in momentum space to the condition
\begin{equation}
        q_i E_{ij} = 0, \hspace{0.3cm} j = x,y,
        \label{eq: rank 2 gauss law}
\end{equation}
which can be easily fulfilled by taking 
\begin{equation}
    \mathbf{E}(\mathbf{q}) = \begin{pmatrix}
        q_y^2 \phi & -q_x q_y \phi \\ -q_x q_y \phi & q_x^2 \phi
    \end{pmatrix}.
\end{equation}  
In this case, the low-energy functional can be written as 
\begin{equation}
    F \simeq \int d^2 q ~\eta_{ij} E_{ij}(\mathbf{q}) E^{ij}(-\mathbf{q}) = \int d^2q  \; \varepsilon(\mathbf{q}) |\phi(\mathbf{q})|^2 
\end{equation}
where the coefficients $\eta_{ij}$ depend on $\alpha$ and $\beta$.

Other contact points located on critical lines in phase diagrams have a local sextic dispersion 
\begin{equation}
    \varepsilon(\mathbf{q}) \simeq \alpha^2 ( q_x^6 + q_y^6) + 2 \beta (q_x^4q_y^2 + q_x^2q_y^4).
\end{equation}
In this case, a symmetric rank three tensor $\mathbf{E}$ can be introduced, following a generalized Gauss law analog to Eq.~(\ref{eq: rank 2 gauss law})
\begin{equation}
    \partial_i E_{ijk} = 0, 
    \label{eq: Gauss law rank 3}
\end{equation}
with $i, j, k$ referring to real space coordinates $x$ and $y$. This Gauss law can be satisfied by taking
\begin{equation}
    \begin{split}
        \mathbf{E} &= \begin{pmatrix}
            E_{xxx} & E_{xxy} & E_{xyx}  &  E_{xyy} \\
            E_{yxx} & E_{yxy} & E_{yyx}  &  E_{yyy}
        \end{pmatrix} \\
        &= \begin{pmatrix}
            q_y^3 \phi & -q_x q_y^2\phi & -q_x q_y^2\phi & q_x^2 q_y\phi \\
            -q_x q_y^2\phi & q_x^2 q_y\phi & q_x^2 q_y\phi & - q_x^3\phi
        \end{pmatrix}.
    \end{split}
\end{equation}
Note this tensor is indeed symmetric since $E_{ijk} = E_{ikj} = E_{jik} = E_{kji}$. The low energy functional can again be expressed using this tensor as
\begin{equation}
    F \simeq \int d^2 q ~\eta_{ijk} E_{ijk}(\mathbf{q}) E^{ijk}(-\mathbf{q}) = \int d^2q  \; \varepsilon(\mathbf{q}) |\phi(\mathbf{q})|^2 
    \label{eq: Energy functional rank 3}
\end{equation}
where coefficients $\eta_{ijk}$ again depend on $\alpha$ and $\beta$. This type of rank 3 tensor seems not to have been previously discussed in the literature \cite{Pretko_2017, Pretko_2_2017, Prem_2018}. A tensor field $\mathbf{E}$ with a functional of the form of Eq.~(\ref{eq: Energy functional rank 3}) and satisfying the Gauss law Eq.~(\ref{eq: Gauss law rank 3}) leads to a structure factor of the form Eq.~(\ref{eq: structure pp rang 3}).

\subsection{Direct link between constraint vector and Gauss Laws}

There exists an alternative way to link the local contact point dispersion to the associated pinch point structure \cite{Yan2023}. Around a contact point $\mathbf{q}_0$ the constraint vector must vanish and thus admits a local expansion 
\begin{equation}
    \mathbf{L}(\mathbf{q_0+q}) = q_i \partial_{q_i} \mathbf{L}(\mathbf{q}_0) + q_i q_j \partial_{q_i}\partial_{q_j} \mathbf{L}(\mathbf{q}_0) + \ldots
\end{equation}
where $\partial_{q_i}\mathbf{L}(\mathbf{q}_0) \neq 0$ for usual quadratic contact points, but becomes zero for contact points with a local dispersion that is at least quartic. Together with the zero temperature constraint Eq.~(\ref{eq: ground state constraint}) it implies a first-order Gauss Law
\begin{equation}
    \partial_{q_i} L_m(\mathbf{q}_0) q_i \mathbf{S}^m_\mathbf{q} =0 
\end{equation}
for a quadratic point, a second-order Gauss Law 
\begin{equation}
    \partial_{q_i}\partial_{q_j} L_m(\mathbf{q}_0) q_i q_j \mathbf{S}^m_\mathbf{q} = 0 
\end{equation}
for a quartic contact point, and more generally a $n$ order Gauss Law for a contact point with a dispersion of order $2n$. This can be obtained directly in real space, expanding the coarse-grained version of the ground state constraint $\mathbf{S}(\mathbf{r}_p) = 0 $ around the plaquette position $\mathbf{r}_p$. To do this, consider the coarse-grained version of spins $\mathbf{S}_m(\mathbf{r})$, defined such that $\mathbf{S}_m(\mathbf{r}_{i,m}) \equiv \mathbf{S}^m_i$ for each lattice site $i$ of sublattice $m$. Next, rewrite it as 
\begin{equation}
    \mathbf{S}_m(\mathbf{r}) = e^{i\mathbf{q}_0\cdot \mathbf{r}  }\bm{\chi}_m(\mathbf{r})
\end{equation}
with $\bm{\chi}_m(\mathbf{r})$ a continuously varying vector field encoding the fluctuations around the contact point configurations. The Taylor expansion of this field for a spin located at $\mathbf{r}_{i,m}$, taken around a plaquette position $\mathbf{r}_p$ is then
\begin{equation}
    \begin{split}
        \bm{\chi}_m(\mathbf{r}_{i,m}) =&\; \bm{\chi}_m(\mathbf{r}_p) + \left[(\mathbf{r}_{i,m}-\mathbf{r}_p) \cdot \bm{\nabla} \right] \bm{\chi}_m(\mathbf{r}_p) \\
        &+ \frac{1}{2} \left[(\mathbf{r}_{i,m}-\mathbf{r}_p) \cdot \bm{\nabla}\right]^2 \bm{\chi}_m(\mathbf{r}_p)+ \ldots
    \end{split}
\end{equation}
Using this expansion the ground state constraint 
\begin{equation}
    \bm{\mathcal{S}}(\mathbf{r}_p) = \sum_m \sum_{i,m \in p} \mathbf{S}_m(\mathbf{r}_{i,m}) = 0
\end{equation}
becomes 
\begin{equation}
     \begin{split}
         L_m(\mathbf{q}_0)  \bm{\chi}_m(\mathbf{r}_p) -  L_m(\mathbf{q}_0) \mathbf{r}_p \cdot \bm{\nabla} \bm{\chi}_m(\mathbf{r}_p) \hspace{0.1 \linewidth} \\
     \hspace{0.1 \linewidth} + \bm{\nabla} \cdot \left[ \sum_{m}\sum_{i,m\in p} \eta_i e^{i\mathbf{q}_0\cdot \mathbf{r}_i } \mathbf{r}_i  \bm{\chi}_m(\mathbf{r}_p) \right] + \ldots = 0
     \end{split}
\end{equation}
It then appears clearly that if a contact point forms at $q_0$, i. e. if the constraint vector $\mathbf{L}(\mathbf{q}_0)$ becomes zero, the fields 
\begin{equation}
    \begin{split}
        \mathbf{E}^\alpha (\mathbf{r}) &\equiv \sum_{m}\sum_{i,m\in p} \eta_i e^{i\mathbf{q}_0\cdot \mathbf{r}_i } \mathbf{r}_i  \chi_m^\alpha(\mathbf{r})\\
        &= \sum_{m} \bm{\nabla}_\mathbf{q} L_m(\mathbf{q}_0) \chi_m^\alpha(\mathbf{r})
    \end{split}
\end{equation}
will obey a Gauss Law $\bm{\nabla}\cdot \mathbf{E}^\alpha$ with $\alpha$ labeling spin components. If the contact point admits a quartic dispersion, meaning $\bm{\nabla}_\mathbf{q} L_m(\mathbf{q}_0) = 0$, this field will become trivial. The relevant tensor field can then be constructed as 
\begin{equation}
    \mathbf{E}_{\mu \nu}(\mathbf{r}_p) \equiv \frac{1}{2} \sum_m\sum_{i,m\in p} \eta_i e^{i\mathbf{q}_0\cdot \mathbf{r}_i } r_i^{\mu}  r_i^{\nu} \bm{\chi}_m(\mathbf{r}_p), \label{eq : Real space rank 2 E}
\end{equation}
which is symmetric by construction and obeys the second order Gauss Laws $\partial_\mu \partial_\nu \mathbf{E}_{\mu \nu} = 0$. The construction of the relevant effective divergentless tensors associated with higher dispersions is similar. The rank of the tensor built will coincide with the pinch point structure for the same reasons as the ones discussed in the previous subsection \cite{Yan_2020, Pretko_2017}.

\subsection{Monte Carlo simulations}

In order to explore the behavior of the systems with temperature, we resorted to Monte Carlo simulations, using the Metropolis algorithm combined with overrelaxion, and lowering the temperature in an annealing scheme. The temperature $T$ is always expressed in units of the coupling with the highest absolute value for each particular set of parameters. We thermalized the system for over 10$^5$ Monte Carlo steps (mcs), and took measurements for twice as many mcs. We worked with systems with  $N=n_0\,L^2$ spins, where $n_0$ is the number of sites in the unit cell ($n_0=2$ for the checkerboard lattice and $n_0=3$ for kagome), and $L$ is the linear size, which we took as $L=24-60$.

In order to look for possible spin-liquid signatures, we measured two quantities: the specific heat per spin ($C_v$), and the static structure factor defined as $S(\mathbf{q})=\frac{1}{N}\sqrt{\left\langle\left|\sum_{j}\mathbf{S}_je^{i\mathbf{q}\cdot\mathbf{r}_j}\right|^2\right\rangle}$. 

On one hand, exotic phenomena such as zero modes or entropic state selection (OBD) may lower the specific heat from its expected equipartition value, which for three-dimensional spins with fixed length is 1 (in units of the Boltzmann constant). In the OBD scenario \cite{OBDVillain}, the specific heat may be lowered by the presence of soft modes with quartic order thermal fluctuations, which are selected from the ground-state manifold since they lower the free energy, as was observed for example in the honeycomb and the kagome lattices \cite{AlbaRosales2016,OBDkagome}. Regarding the zero modes, which do not contribute to the specific heat, for systems where the Hamiltonian can be rewritten as the sum of plaquettes, the number of zero modes in the model may be expressed as \cite{SLPiro2,Rehn_Moessner_2016,AlbaPujol2018,AlbaRosales2021}:
\begin{equation} \label{eq:zero-modes}
 F=\frac{q}{b}(n-1)-n  
\end{equation}
where $q$ is the number of spins per plaquette, $b$ is the number of plaquettes that share the same spin, and $n$ is the dimension of the spin, which in this work is $n=3$. The specific heat at low temperatures is reduced by these zero modes to: 
\begin{equation} \label{eq:cv}
C_v=b\,n/2q
\end{equation}
As we see, in these two expressions only enter the ratio $b/q$. This is an important point for the three cases that we study here, indeed, in building the generalized plaquettes, the coefficients $q$ and $b$ increase, but the ratio is always kept constant. 

On the other hand, an inspection of the structure factor in reciprocal space may reveal features such as the above-mentioned pinch points. Furthermore, the structure factor lends itself well to comparison with both LTA analysis and experimental findings, enabling the generation of explicit predictions for  Neutron Scattering results.

\section{Example 1: the checkerboard lattice}
\label{sec: Checkerboard}

\begin{figure}[h!]
	\centering \includegraphics[width=\linewidth]{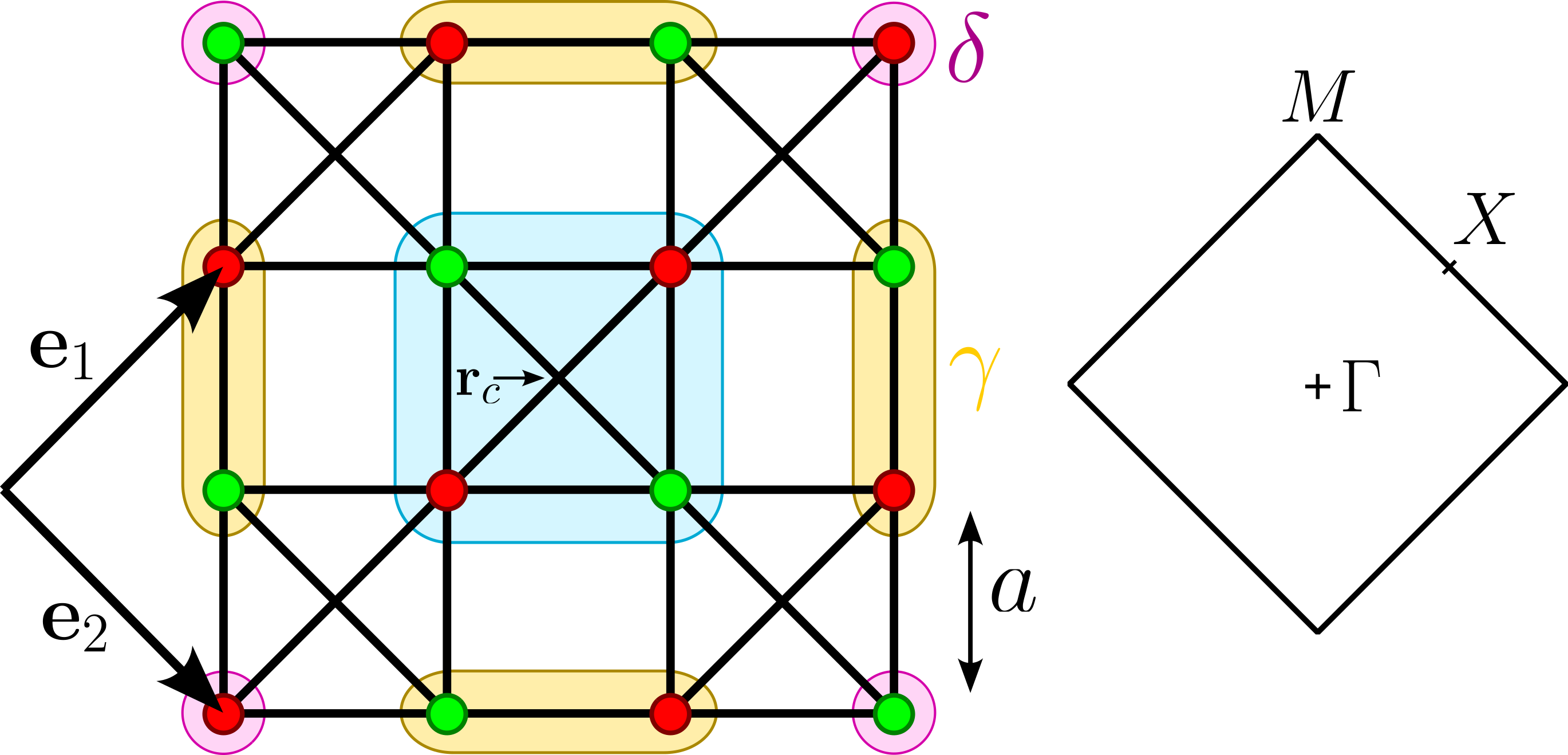}
        \caption{Extended plaquette for the checkerboard lattice. The four spins from the standard checkerboard plaquette come in the total spin definition (Eq.~(\ref{eq: Spin tot Ckb})) with coefficient $1$, and are surrounded  by a square. The pairs of side spins have  coefficient $\gamma$ and corner points, coefficient $\delta$. Colors in the sites of the lattice indicate different sublattices. The primitive lattice vectors are denoted by $\mathbf{e}_1$ and $\mathbf{e}_2$, and the corresponding Brillouin zone is given on the right of the figure.}
        \label{fig: checkerboard2}
\end{figure}

As a first example, we examine the checkerboard lattice and define the effective spin $\mathcal{S}_p$ of an extended plaquette $p$ as follows:
\begin{equation}
    \bm{\mathcal{S}}_p = \sum_{i \in p } \mathbf{S}_i
 + \gamma \sum_{i \in \langle p \rangle } \mathbf{S}_i + \delta \sum_{i \in \langle\langle p \rangle \rangle } \mathbf{S}_i, \label{eq: Spin tot Ckb}
 \end{equation}
 where the first sum corresponds to the spins located at the vertices of a crossed square denoted by $p$, as illustrated in Fig.~\ref{fig: checkerboard2}. The second and third sums involve the sites adjacent to the crossed square, with the first sum considering the sites connected by $\gamma$, and the second sum considering the sites connected by $\delta$. We propose to explore this model using first the LTA to obtain an approximate low-energy description of the system. We will then analyze the constraint vector $\mathbf{L}_\mathbf{q}$ and its topological properties, which will provide further insights into the system's behavior. Finally, we will perform Monte Carlo simulations to verify the predictions made by the LTA and investigate the model's properties in more detail. Additionally, we will describe the behavior of the system using a real space gauge field description, which can provide a useful framework for understanding the underlying physics.

%

\subsection{Luttinger-Tisza approximation}

\begin{figure}
    \centering
    \includegraphics[width = \linewidth]{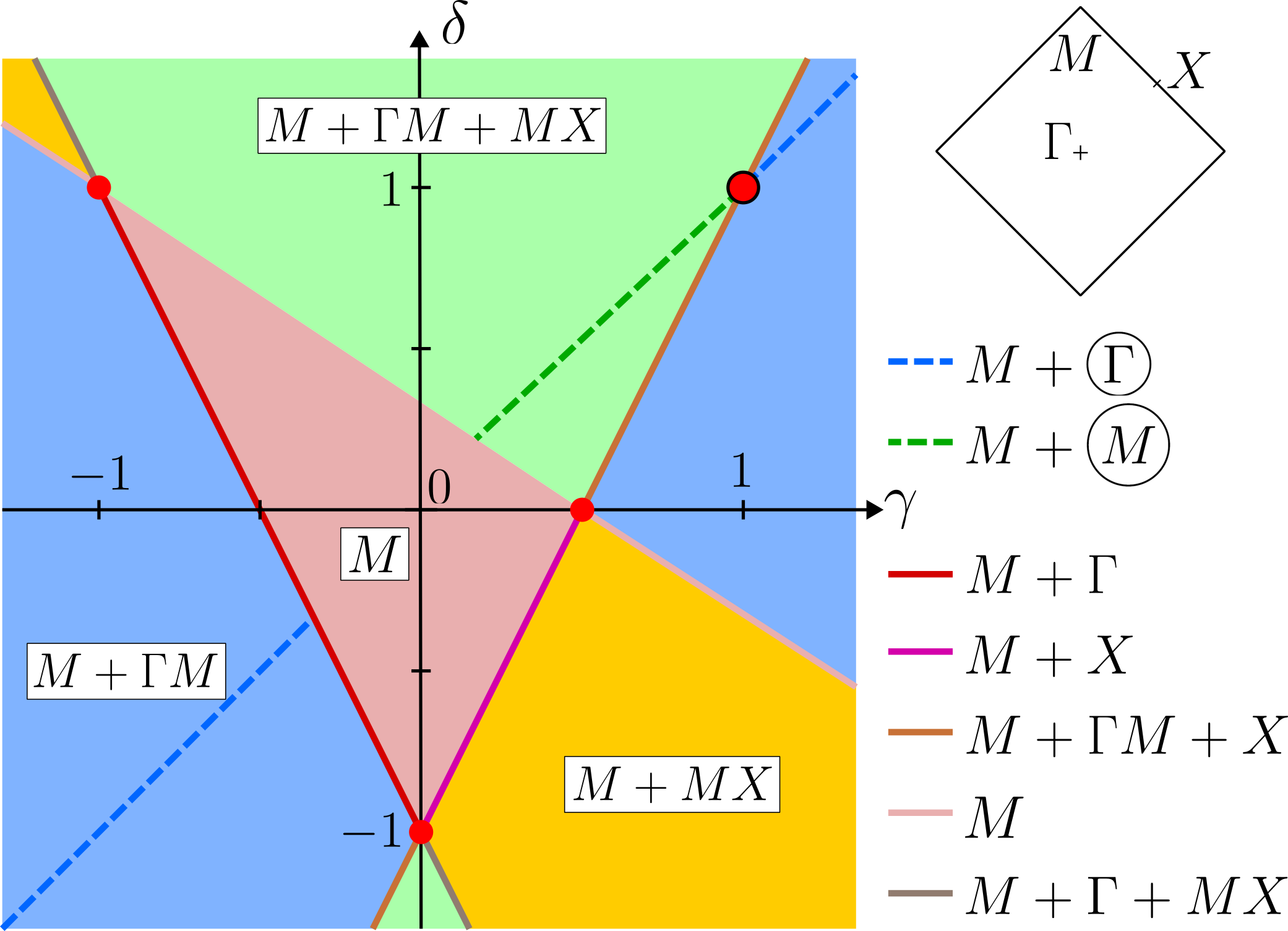}
    \caption{Phase diagram, with $\gamma$ in abscissa and $\delta$ on the ordinate. The labels indicate the position of the pinch points. There are four major phases, separated by critical lines with distinct features (see text for details). There are four special points marked in red, which show particularly special features. The phases corresponding to three of the four special points marked with red dots are given in fig \ref{fig: ckb pinch lines}. The last one, located at point $\gamma = 0$, $\delta = -1$ corresponds to the phase $M + \Gamma + X$. On the top right corner the position of the  special points $\Gamma, M, X$ along the BZ is given.}
    \label{fig: ckb pahse diagram}
\end{figure}

As announced in the previous section, the first step is to construct the constraint vector function. Since the lattice has two sites in the unit cell, $\mathbf{L_q}$ is a two-component vector. Moreover, the inversion symmetry with respect to the center of the plaquette ($\mathbf{r}_c$ in Fig.~\ref{fig: checkerboard2}) ensures that its components are real. The constraint vector writes explicitly
\begin{equation}
	\begin{split}
			\mathbf{L}(\mathbf{q}) = 
                2\begin{pmatrix}
			         \cos \left(\frac{1}{2}\mathbf{e}_1\cdot \mathbf{q} \right)   \\ 
			         \cos \left(\frac{1}{2}\mathbf{e}_2\cdot \mathbf{q} \right)
		        \end{pmatrix} 
            + 2\delta \begin{pmatrix}
		                  \cos \left(\frac{3}{2}\mathbf{e}_1\cdot \mathbf{q} \right) \\
		                  \cos \left(\frac{3}{2}\mathbf{e}_2\cdot \mathbf{q} \right) 
	               \end{pmatrix}\\ 
        + 2\gamma \begin{pmatrix}
			\cos\left(\left(\mathbf{e}_1 + \frac{1}{2}\mathbf{e}_2\right)\cdot \mathbf{q} \right) + \cos\left(\left(\mathbf{e}_1 - \frac{1}{2}\mathbf{e}_2\right)\cdot \mathbf{q} \right)  \\
			\cos\left( \left( \frac{1}{2}\mathbf{e}_1+\mathbf{e}_2 \right)\cdot \mathbf{q} \right) + 	\cos\left( \left( \frac{1}{2} \mathbf{e}_1-\mathbf{e}_2 \right) \cdot \mathbf{q} \right)  
		\end{pmatrix} 
	\end{split}
    \label{L checkerboard}
\end{equation} 
with $\mathbf{e}_1 = a(1,1)^t$ and $\mathbf{e}_2 = a(1,-1)^t$ as depicted on Fig.~\ref{fig: checkerboard2}.
These properties imply that there is a unique flat band associated with a dispersive band, which is simply proportional to the norm of the constraint vector. The dispersive band always touches the first band at different points of the Brillouin zone (BZ), but does so in different ways depending on the values of the parameters $\gamma$ and $\delta$. The contact surface between the two bands can take the form of isolated points, corresponding to pinch points as depicted in the previous section. The shape of the band around the contact point then determines the type of pinch point, see Eqs.~(\ref{eq: quadratic pinch point}) and (\ref{eq: quartic pinch point}). The contact surface can also manifest as a contact line, which occurs when the values of $\gamma$ and $\delta$ result in $\mathbf{L_q}$ is always zero,  defining a closed curve in reciprocal space with the equation $q_x = f(q_y)$\cite{Benton2016}. These lines do not correspond to pinch points but to degeneracy lines and the reason for their presence is discussed in subsection \ref{subsec :  Degeneracy lines CkB}. The results given by the LTA analysis are summarized in the phase diagram presented in Fig.~\ref{fig: ckb pahse diagram}.  The phases are labeled using the standard high-symmetry points in the BZ: $\Gamma$, $M$, and $X$.  For example, $M + MX$ corresponds to a situation where there is a contact point located on each equivalents corner point $M$, and another on each equivalent axis linking a point $M$ to a point $X$. The notation $M + \encircled{\Gamma}$ indicates a contact point located on the corner point $M$, and a contact line encircling the central point $\Gamma$. 

Four distinct extended phases, each presenting only pinch points, have been identified in this study. The first phase is characterized by pinch points only at the $M$ points, while the second and third phases have four additional contact points arising in the $\Gamma M$ and $XM$ axes, respectively. In the fourth phase, pinch points appear in both the $\Gamma M$ and $XM$ points, resulting in eight additional contact points with the lowest energy band. These extended phases are separated by critical lines, represented with solid lines in the phase diagram,  which always correspond to the emergence of new pinch points. On one side of the critical line, the pinch point disappears, while on the other side, it splits into new pinch points. For instance, consider the line $M + \Gamma$ separating the $M$ and $M + \Gamma M$ phases. As one approaches this line from the $M$ phase side, a pinch point emerges at point $\Gamma$ and then splits itself into four pinch points located on each segment $\Gamma M$. When moving away from this critical line, the new pinch points migrate from point $\Gamma$ towards point $M$. This suggests that some pinch points along these critical lines could be higher-order pinch points since they are split into sub-pinch points. This issue is discussed in more detail below.

\begin{figure}
    \centering
    \includegraphics[width=\linewidth]{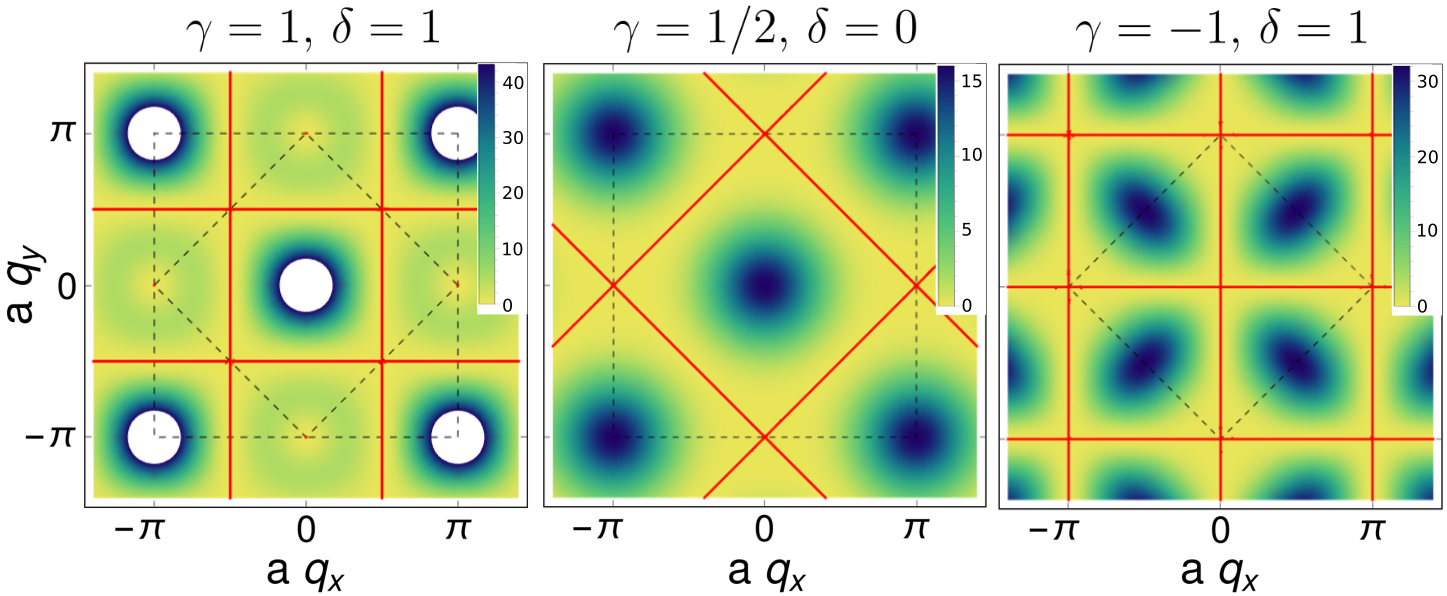}
    \caption{Heat-maps of the excited band $\varepsilon(\mathbf{q})$ obtained with the LTA for the checkerboard lattice for three cases where there are contact lines between the lowest-energy flat bands and the excited one (degeneracy lines). Red lines correspond to the equipotential lines $\varepsilon(\mathbf{q}) = 0$. }
    \label{fig: ckb pinch lines}
\end{figure}

The special line associated with equation $\gamma = \delta$, represented by a dotted line in the phase diagram, is different in essence since it does not separate two distinct phases. It hosts two kinds of critical states where circular contact lines are observed around point $\Gamma$ or $M$. Finally, there are four critical points located at the intersection of the critical lines. Three of these special points show straight contact lines in the LTA spectrum, located in different positions (see Fig. \ref{fig: ckb pinch lines}), a phenomenon that is also discussed below. The fourth point, located at $\gamma = 0$, $\delta = -1$ corresponds to the phase $M + \Gamma + X$, likely to be the only phase holding two different types of higher-rank pinch points.

Note that, for all the values of the parameters, there are always contact points between the two bands at the $M$ points (as in the $\gamma=\delta=0$ case), implying that the present model is always an algebraic spin-liquid. It is then natural to search for an associated Gauss law, as we explain below.
As mentioned above, the fact that there exist pinch points that split into sub-pinch points suggests that these pinch points should be higher-order pinch points. There are actually three different situations corresponding to the three critical lines separating extended phases. 

The line of equation $\delta = -1 + 2\gamma$ presents pinch points located at $X$ points that split into two sub-pinch points located on the BZ boundaries. The local dispersion around these special pinch points is quartic along the BZ boundary but quadratic along the perpendicular direction $X-\Gamma$. This Lifshitz type of behavior \cite{Hornreich_1975} corresponds to a special pinch point in the structure factor, as we see in Fig.~\ref{fig:SqCheck}.

The critical line observed for $\delta = -1 - 2\gamma $ possesses a contact point located at $\Gamma$, which splits into four sub-pinch points along axis $\Gamma M$. These special pinch points have a quartic dispersion.
As pointed out in Sec.~ \ref{subsec: Low-temperature emergent Gauss tensors}, this corresponds to the existence of an underlying gauge field theory using a rank-two tensor $E_{ij}^{\alpha}$ for each spin component $\alpha$, satisfying a generalized Gauss law $\partial_i E_{ij}^\alpha =0$ for each spin component $\alpha$ \cite{Pretko_2017, Pretko_2_2017, Yan_2020, Prem_2018}. This corresponds in the structure factor to pinch points presenting a four-fold degeneracy.

The last critical line of equation $\delta = (1-2\gamma)/3$ possesses a high-rank pinch point located at point $M$ which splits into eight sub-pinch points along axis $\Gamma M$ and $MX$. This pinch point shows a sextic dispersion and thus corresponds to an emerging rank-three tensor satisfying a generalized Gauss law $\partial_i E_{ijk}^\alpha =0$ for each spin component $\alpha$, as presented in Sec.~\ref{subsec: Low-temperature emergent Gauss tensors}.

\subsection{Topological properties of the Constraint Vector function}

As stated above, the vector $\mathbf{L}_\mathbf{q}=(L_{1,\mathbf{q}},L_{2,\mathbf{q}})$ is a two-component real vector. The topological defects of such a vector field are vortices, for which the field norm $\|\mathbf{L}_\mathbf{q} \|$ must vanish at their center, implying the existence of pinch points\cite{Benton_Moessner_2021}. In principle, the positions, number, and topological indexes of the vortices may be another equivalent way to classify the phases discussed above. 

To detect the vortices, we compute the associated vorticity, defined as 

\begin{equation}
Q=\frac{1}{2\pi}\oint_{C} \,d\textbf{q}\cdot(\tilde{L}_1\nabla_{\textbf{q}}\tilde{L}_2 - \tilde{L}_2\nabla_{\textbf{q}}\tilde{L}_1)
\label{eq:int_vortex}
\end{equation}
where the integral in Eq.~(\ref{eq:int_vortex}) is defined for a closed path $C$ and $\tilde{L}_1$ and $\tilde{L}_2$ are the components of the normalized vector ${\bf \tilde{L} = L_q/\|L_q\|}$. This calculation, although simple in principle, requires some attention, especially in the definition of the closed path $C$ in phases with multiple pinch points that are not positioned in special points in the BZ. We chose circles surrounding these points, with a radius smaller than the distance to other singularities, to avoid subtracting or adding vorticities.

\begin{figure*}[htb!]
        \centering \includegraphics[width=1.0\textwidth]{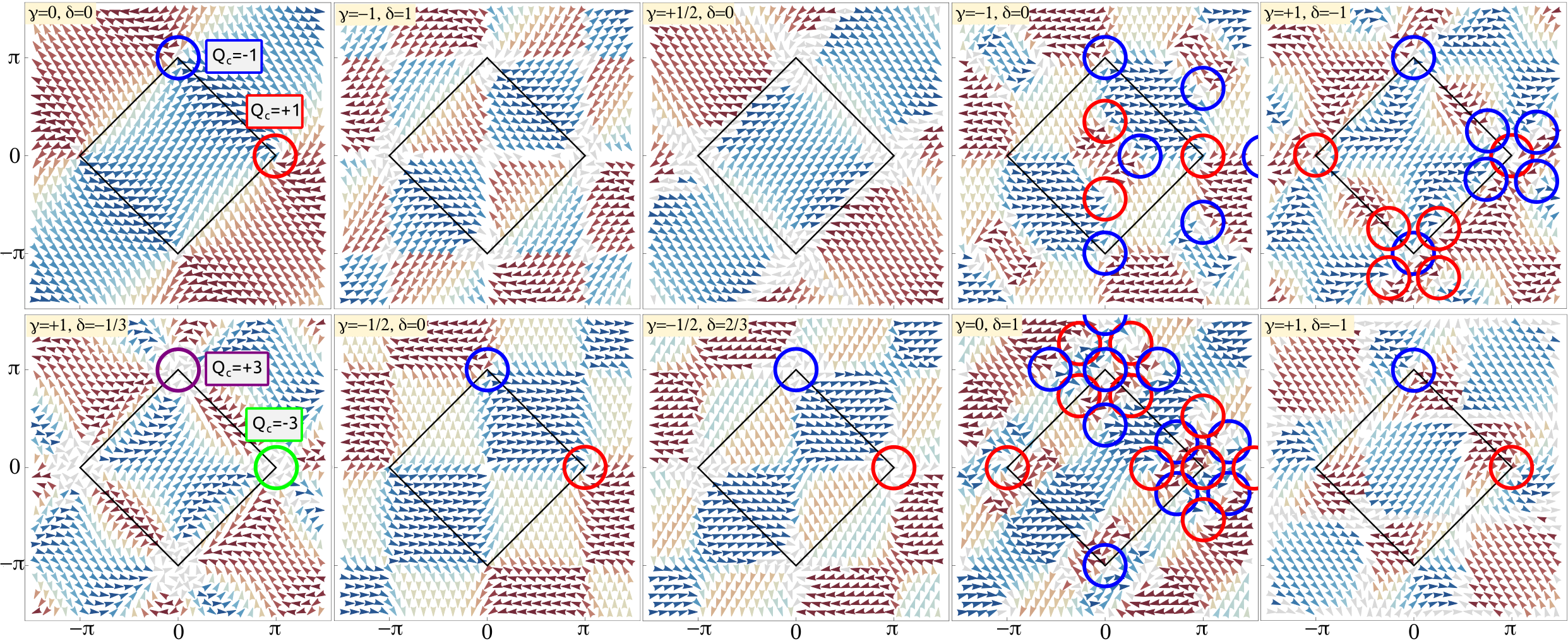}
        \caption{Checkerboard case: Vector plots of the constraint vector $\mathbf{L}_\mathbf{q}$  for different $\gamma, \delta$ parameters in the extended checkerboard lattice. Circles mark the position of the vortices/antivortices, and the value of the corresponding winding number is indicated.}
        \label{fig: checkerboard-vectorplot}
\end{figure*}

In Fig.~\ref{fig: checkerboard-vectorplot} we show the vector plots of the constraint vector $\mathbf{L}_\mathbf{q}$ for different values of $(\gamma, \delta)$, where in general clear vortices can be seen centered in the position of the pinch points. In most cases, we obtain that the pinch points are associated with pairs of vortices-antivortices with vorticity $|Q|=1$.  Nonetheless, there are a few special cases to mention. First, we see for example that  for $(1,-1/3)$, the vortices at the $M$ points have a higher topological charge, $|Q|=3$. For these parameters, as was discussed above, higher-order pinch points are expected, and thus we might expect higher vorticity. However since the vorticity is a quantity that affects the long-range structure, one expects that the total vorticity of a region including multiple vortices should be conserved when different pinch points merge into a single higher-order pinch point. This is indeed the case, looking at the line $\delta=-1+2\gamma$, the quartic pinch point located at $\Gamma$ results from the merging of 2 pairs of vortex-antivortex. It is then not surprising that the resulting pinch point has  zero vorticity, see for example the point $(-0.5,0)$ in Fig. \ref{fig: checkerboard-vectorplot}. If we now look at the critical line separating the phases $M+ \Gamma M$ and $M+MX$ the situation is different, since on the $M+ \Gamma M$ side a vortex of charge $1$ located at point $M$ is surrounded by four charges $-1$ antivortices on the axis $\Gamma M$ (see point $(-1,0)$ on Fig.~\ref{fig: checkerboard-vectorplot}). When merging, these five vortices will thus lead to the formation of an antivortex of vorticity $-3$, see Fig. \ref{fig: checkerboard-vectorplot}. On the  $M+MX$ side of the line, this pinch point splits again into four antivortices of vorticity $-1$ located on $XM$ axis, and one charge $1$ vortex located at $M$ (see point $(1,-1)$ on Fig.~\ref{fig: checkerboard-vectorplot}).

This illustrates why the topological analysis does not allow spotting some pinch points if they result from the merging of vortex-antivortex pairs.

Note that for the special points $(\gamma,\delta)=(0.5,0)$ and $(-1,1)$, there are degeneracy lines that reach the point $M$ at which the vector $\mathbf{L_q}$ vanish. For these cases, the computation of the vorticity is thus ill-defined.

\subsection{Monte Carlo simulations and thermal effects}
\label{subsec:MC-checkerboard}

The checkerboard lattice, which is the ``simple'' plaquette from Fig.~\ref{fig: checkerboard2}, setting $\gamma=\delta=0$ in Eq.~(\ref{eq: Spin tot Ckb}), may be interpreted as corner sharing squares where all four vertices of the plaquette are connected. Since each spin is shared for two plaquettes, replacing $q=4$ and $b=2$ in Eq.~(\ref{eq:zero-modes}), we see that there is one zero mode and thus the specific heat, following Eq.~(\ref{eq:cv}) is expected to be $3/4$ at lower temperatures if there are no additional soft modes. This is in fact also the case for the pyrochlore lattice \cite{SLPiro1,SLPiro2}. 
When $\gamma$ is turned on, the coefficients $q$ and $b$ jump to respectively the values of $12$ and $6$, and if $\delta$ is also turned on, they become respectively  $16$ and $8$. The crucial point to note is that the ratio of $b$ to $q$, which determines the value of the specific heat, is always maintained at $1/2$.

\begin{figure}
    \centering
    \includegraphics[width=0.9\columnwidth]{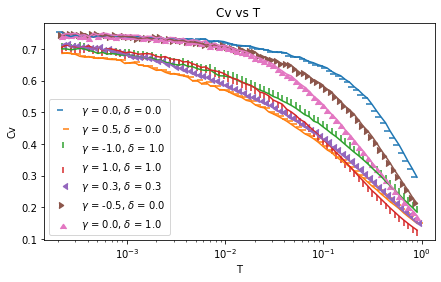}
    \caption{Specific heat per spin as a function of temperature for the extended checkerboard lattice for different parameters $(\gamma,\delta)$ obtained from MC simulations. The temperature is in units of the highest effective coupling.}
    \label{fig:cv_check}
\end{figure}

\begin{figure*}
    \centering
    \includegraphics[width=1\textwidth]{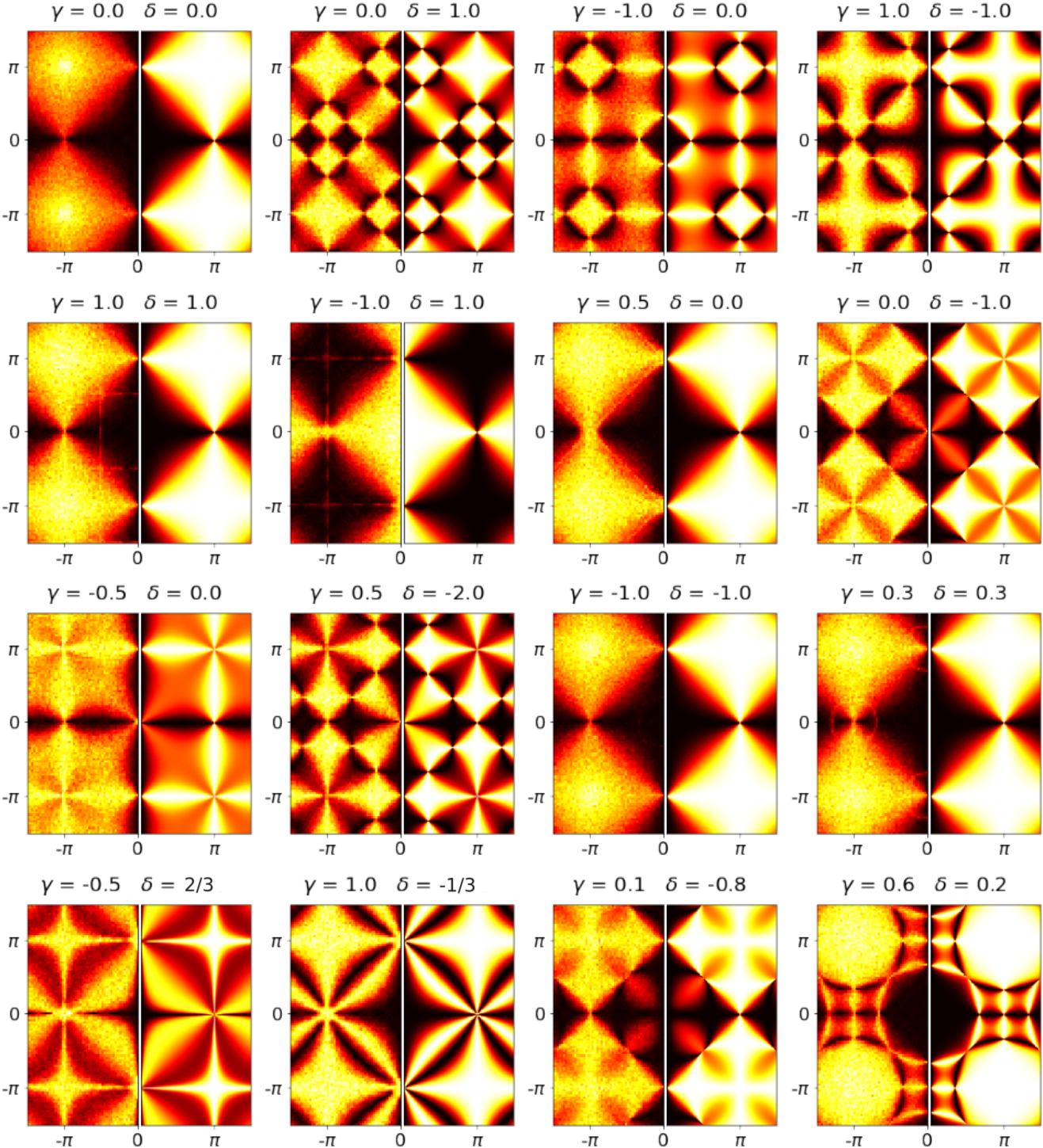}
    \caption{Comparison of $S_q$ obtained from MC simulations (at $T=0.0002$, left side of each panel) and Projective Analysis results (right) for different regions of the LTA phase diagram (Fig.~\ref{fig: ckb pahse diagram}). First row: representative points for the  four extended regions. Second Row: points indicated in red. Third and Fourth: example points at different degenerate lines. Note that for $\gamma = \delta =-1$, when looking carefully, the structure factor obtained through MC simulations presents a circular degeneracy line encircling the  $\Gamma$ point, matching with LTA predictions. }
    \label{fig:SqCheck}
\end{figure*}

The $C_v$ as a function of temperature and the low-T structure factor are shown in Figs. \ref{fig:cv_check} and in the first panel of Fig. \ref{fig:SqCheck}.
MC simulations for the extended model show an excellent agreement with the analytical predictions, as seen in the structure factors $S(\mathbf{q})$ plots in Fig.~\ref{fig:SqCheck} for different parameter sets in the phase diagram. Regarding the effect of temperature, we see that the zero modes remain present for all $(\gamma,\delta)$ values at low temperature, pinch points remain in the structure factors, and the $C_v$ is lowered from $3/4$ at some special points. We show this in Fig.~\ref{fig:cv_check}, where we compare the specific heat of some typical cases ($(\gamma, \delta)=(0,0),(-0.5,0),(0,1)$) with the particular values $(1,1),(-1,1),(0.5,0)$, where the $C_v$ remains lower than $3/4$ at the lowest simulated temperature. This agrees with the LTA prediction since in these cases there are additional degenerate lines, which are related to states with particular sub-extensive degeneracy, as it is described in subsection \ref{subsec :  Degeneracy lines CkB}. This degeneracy is also seen as additional lines in the structure factor, as shown in Fig.~\ref{fig:SqCheck}. Notice that for $(-1,1)$ the degeneracy lines are seen in the EBZ, and not in the first BZ. In the LTA, contact lines between the second band and the lowest energy flat band were also found as circles for a region in the $\gamma=\delta$ line. As an example, the specific heat for $(0.3,0.3)$ is  lower than $3/4$ at least up to the lowest simulated temperature,  indicating that this additional semi-extensive degeneracy  introduces soft modes.

Moreover, the structure factors in Fig.~\ref{fig:SqCheck} also show the presence of higher order pinch points:   four-fold pinch points at $\Gamma=0$ at the $\delta=-1-2\gamma$ line,  illustrated for $(-0.5,0)$, six-fold pinch points at the $\delta=(1-2\gamma)/3$ line, shown for $(-0.5,2/3)$ and $(1,-1/3)$, and Lifshitz  pinch points  in the $\delta=-1+2\gamma$ line,  exemplified for $(0.1,-0.8)$ and $(0.6,0.2)$. In the special case $(0,-1)$ there are two types of pinch points: quartic ones at the $\Gamma$  point and Lifshitz ones in the $X$ points of the BZ.

Figure \ref{fig:SqCheck} compares $S(\mathbf{q})$ obtained from a Monte Carlo simulations (left half) with the results from the Projective Analysis\cite{Henley2005} (right half)
\begin{equation}
    S(\mathbf{q}) = \sum_{m,n} \langle\mathbf{S}^m_{-\mathbf{q}}\cdot\mathbf{S}^n_{\mathbf{q}} \rangle\propto 1 - \frac{L_1(\mathbf{q}) L_2(\mathbf{q})}{\|\mathbf{L}\|^2}
\end{equation}

\noindent obtained directly from  Eq.~(\ref{eq: projector unique real Lq}).
The comparison shows a strong agreement between the two approaches, confirming that this analytical method allows identifying both the location and the structure of pinch points. This method fails however to reveal the existence of degeneracy lines, the limit of the structure factor $S(\mathbf{q})$ being well defined when approaching a degeneracy line, even if the denominator $\|\mathbf{L}\|^2$ vanishes.  

%
\subsection{Gauss Law}
\label{subsec: Gauss law checkerboard}

Here we demonstrate the presence of an effective $U(1)$ gauge field theory allowing for dipolar correlations related to a photon-like propagator\cite{Henley2005} producing pinch points in the structure factors. This effective gauge theory is based on the presence of a divergence-free gauge field. 

\begin{figure}
    \centering
    \includegraphics[width = 0.95\linewidth]{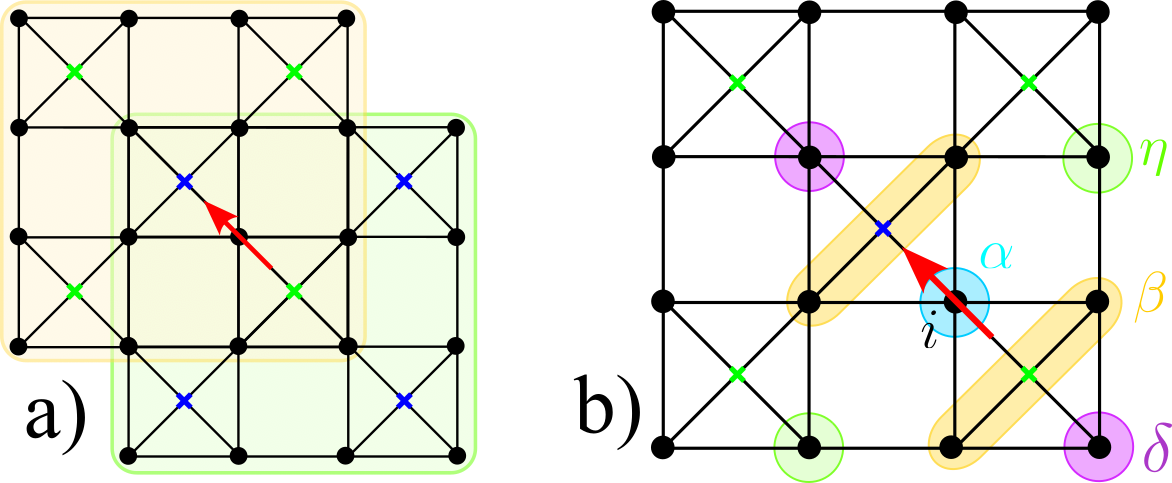}
    \caption{a) The intersection of two neighboring extended plaquettes. The co-dual lattice, which is a square lattice, is bipartite. The sites are then of two types, depicted with green and blue crosses. Bonds can be oriented from the green sites toward the blue ones. Flux attached the oriented bond linked to the two plaquettes shown, depicted in red, can only be made of spins belonging to both plaquettes. b) Scheme construction for the Gauss Law analysis in the extended checkerboard lattice model. The flux associated with the bond $i$, oriented in the red arrow direction, gets contributions from the surrounding highlighted spins with coefficients $\alpha$, $\beta$, $\eta$, and $\delta$.
    See the text for details.} 
    \label{fig: Ckb GL}
\end{figure}

To construct this field, we consider the lattice with vertices located at the centers of each original crossed plaquette. The resulting lattice is simply a square lattice and is thus bipartite, meaning that each bond can be ``oriented'' from, say, sublattice $A$ to sublattice $B$. The next step is to associate a flux on these oriented bonds such that the sum of the assigned fluxes incoming in one vertex is equal to the total spin $\bm{\mathcal{S}}_p$ of this plaquette, which must be zero for any ground state configuration. In this way, the constructed field will be divergent-free, as requested. The question is now to understand how to constitute the fluxes attached to each bond. By symmetry, the flux through each bond can be expressed as a linear combination of the nine spins associated with the intersection of the two extended plaquettes linked by that bond, see Fig.~\ref{fig: Ckb GL}.a). Taking the notation from Fig.~\ref{fig: Ckb GL}.b), the flux associated with the  bond labeled by the index $i$ and oriented along the direction of the red arrow can be defined as
\begin{equation}
        \bm{\Pi}_i = \alpha\,\mathbf{S}_i + \beta \sum_{j\in \text{yellow} }\mathbf{S}_j + \eta \sum_{j \in \text{green}} \mathbf{S}_j + \delta \sum_{j \in \text{pink}} \mathbf{S}_j.
\label{param}
\end{equation}
where $\alpha$, $\beta$ and $\eta$ are real coefficients. For the sum of four incoming fluxes to be equal to the total spin $\bm{\mathcal{S}}_p$ the weight coefficients $\alpha$, $\beta$, and $\eta$ must satisfy the constraints
\begin{equation}
    \begin{split}
    &\alpha + 2\beta + \delta = 1, \\
    &\beta + \eta = \gamma.
    \end{split}
\end{equation}
These two conditions can always be satisfied for any value of parameters $\gamma$ and $\delta$. This means that one can always build the fluxes  $\bm{\Pi}_i$ with zero divergence on each vertex whatever the values of the parameters $\gamma$ and $\delta$ are. Moreover, it turns out that different choices for the parameters in Eq.~(\ref{param}) produce flux configurations that differ by loops of the shorter possible length (elementary squares in the present case) and thus do not change the large scale behavior of the system, nor its coarse-grained effective action that we describe below. This  remark is also valid for the two other examples, defined on the kagome lattice (sections \ref{sec:Gauss-Law-2} and \ref{sec:Gauss-Law-3}), that we describe in this work. 

Following Henley \cite{Henley2005}, one can then build a rank-two tensor, called polarization tensor $P_{j}^\alpha(\mathbf{r}_i)$, defined on each oriented bond $i$ located at the position $\mathbf{r}_i$ as
\begin{equation}
    P_{j}^\alpha(\mathbf{r}_i) = \Pi_i^\alpha u_i^j
    \label{eq: Polarization tensor definition}
\end{equation}
where $\alpha=x,y,z$ is the index of the spin components. The vector $\mathbf{u}_i$ is the vector giving the direction and the orientation of the link $i$ considered. The coarse-grained version of this tensor field then satisfies, by construction, the three Gauss equations
\begin{equation}
    \bm{\nabla}\cdot \mathbf{P}^\alpha = 0, \label{eq: Gauss condition}
\end{equation}
one for each spin component $\alpha$. \\

The simplest zero temperature energy functional has the usual structure 
\begin{equation}
    F_{\text{tot}}\left({\mathbf{P}}\right) = \frac{\kappa}{2} \int d^2 \mathbf{r} \sum_{i,\alpha} \left(P_{i}^\alpha\right)^2, \label{eq: free energy T0}
\end{equation}
leading, with the Gauss condition (\ref{eq: Gauss condition}), to the existence of pinch points in the structure factor \cite{Henley2005, Prem_2018}. For higher temperatures, consideration of symmetry arguments \cite{Henley2005, Prem_2018} suggests that entropy should favor configuration with a small polarization tensor norm. This indeed corresponds to the possible existence of small fluxes loops, that can be flipped without an energy cost. Since entropy favors the appearance of such loops, it will consequently favor configurations with small polarization tensors. The total free energy should thus behave, considering that entropy effects dominate ($F \sim -TS$), as 
\begin{equation}
    \frac{1}{T}F_{\text{tot}}\left({\mathbf{P}}\right) = \frac{\kappa'}{2} \int d^2 \mathbf{r} \sum_{i,\alpha} \left(P_{i}^\alpha\right)^2. \label{eq: free energy}
\end{equation}
This functional has exactly the same structure as the zero-temperature one. This implies in both cases that the polarization tensor is analog to a magnetic field, meaning that correlations functions do have the longitudinal fluctuations projected out and take the form 
\begin{equation}
    \langle P_i^\alpha (-\mathbf{q}) P_j^\beta (\mathbf{q}) \rangle \propto \delta_{\alpha \beta} \left( \delta_{ij} - \frac{q_i q_j}{\|\mathbf{q}\|^2} \right)
\end{equation}
in momentum space. This corresponds to a function having different limits when $\|\mathbf{q}\| \to 0$, forming the pinch points. 

\subsection{Higher order pinch points}
\begin{figure}
    \centering
    \includegraphics[width = \linewidth]{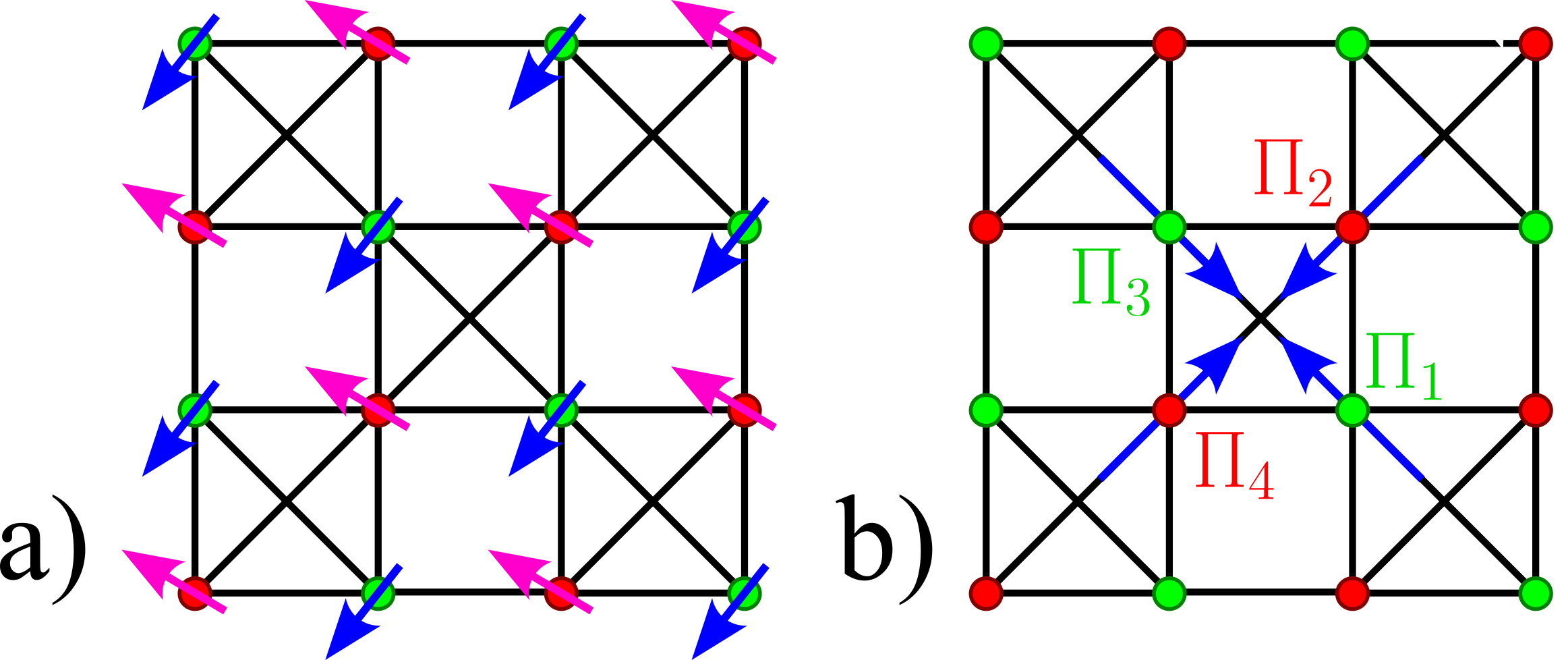}
    \caption{a) Real space spin configurations associated with the existence of a contact point located in $\Gamma$, thus corresponding to a wave vector $\mathbf{q}_0 = 0$. These configurations repeat the same spin all along the system for each sublattice. b) Representation of the fluxes $\bm{\Pi}_i$ placed on the bonds linked to an extended plaquette. These fluxes are attached to the underlying bonds, which can be oriented as depicted with blue arrows.   }
    \label{fig: Ckb G configs and fluxes}
\end{figure}

The fluxes construction above appears to be valid for every value of the parameters $\gamma$ and  $\delta$. The coarse-grained version of the associated polarization tensor (Eq. (\ref{eq: Polarization tensor definition})) obeys a first-order Gauss Law and thus implies, as depicted above, the existence of two arms pinch points. There are however critical lines in the phase diagram presenting more complex pinch points, which seems in contradiction with the above construction. The point is that for specific configurations associated with certain pinch points locations, the coarse-grained version of the associated polarization tensor (Eq. (\ref{eq: Polarization tensor definition})) can become identically zero. This can be illustrated with the case of a contact point located in $\Gamma$, corresponding to four-fold pinch points. The configurations associated with such a contact point correspond to a 0 wave vector, and thus repeat a unique spin on each sublattice, as depicted on Fig.~\ref{fig: Ckb G configs and fluxes} a). For these specific configurations, the fluxes built following the previous subsection appear to be equal if they are attached to opposed bonds. On Fig.~\ref{fig: Ckb G configs and fluxes} b) this corresponds to have $\bm{\Pi}_1 = \bm{\Pi}_3$ and $\bm{\Pi}_2 = \bm{\Pi}_4$. This means that taking the coarse-grained version of the polarization tensor (\ref{eq: Polarization tensor definition}) will imply to sum oriented fluxes of equal magnitude and opposed directions, producing a field that is zero on each plaquette center $\mathbf{r}_p$. In this situation, by analogy with Eq.~(\ref{eq : Real space rank 2 E}), an infinitesimal rank two polarization tensor can be defined as
\begin{equation}
    \mathbf{E}^{ij}_{l} = u_l^i u_l^j \bm{\Pi_l}, \label{eq def P rank 2}
\end{equation}
on each lattice bond $l$. The coarse-grained version of this polarization tensor, symmetric by construction, will satisfy a generalized Gauss Law $\partial_i \partial_j \mathbf{E}^{ij}(\mathbf{r}_p) = 0$ at each plaquette center $\mathbf{r}_p$. This comes from the fact that writing this double divergence as a lattice derivative leads to the relation 
\begin{equation}
    \partial_i \partial_j \mathbf{E}^{ij}(\mathbf{r}_p) \simeq \sum_{l \to p} \bm{\Pi}_l, 
\end{equation}
with the sum being over the bonds $l$ connected to the plaquette center located in $\mathbf{r}_p$ among the co-dual lattice. The above relation thus implies a Gauss law because we built the fluxes in such a way that 
\begin{equation}
    \sum_{l \to p} \bm{\Pi}_l = \bm{\mathcal{S}}_p = 0.
\end{equation}
Consequently, it appears that the construction of geometrical fluxes carried out in the previous subsection can be in fact also associated with the tensor of rank superior to $1$, allowing us to explain why there are pinch points in the phase diagram presenting more than two arms. \\

The existence of an underlying polarization tensor provides an explanation for the presence of pinch points for all values of $\gamma$ and $\delta$. However, this alone does not account for the observation of contact lines in the LTA results, which manifest as bright lines in the MC simulations. Consequently, in the following section, we put forward an explanation for the occurrence of these degenerate lines.

\subsection{Degeneracy lines}
\label{subsec :  Degeneracy lines CkB}

The combination of LTA analysis and MC simulations presented above demonstrate that certain special lines emerge in momentum space, corresponding to a linear band contact $\varepsilon(\mathbf{q}) = 0$ within the context of LTA. As detailed in subsection \ref{subsec:MC-checkerboard}, the MC simulations confirm that these lines give rise to strong spin correlations. If we consider for example the case $\delta = \gamma = 1$, we notice that the structure factor is similar to the ones observed for slightly different parameter values, but with the addition of bright lines (see Fig.~\ref{fig:SqCheck}). These lines can be seen as the addition of terms of the form 
\begin{equation}
    S_{dl}(\mathbf{q}) = \delta(c+q_i), \hspace{1cm} i = x,y \label{eq : lines S(q)}
\end{equation}
with $c$ the offset giving the positions of the lines in momentum space. This can be understood by looking at special real space configurations. Let us start with the simple case $\gamma = \delta = 1$. For these special values of the parameters, the total spin of an extended plaquette $p$ is simply 
\begin{equation}
    \bm{\mathcal{S}}_p = \sum_{i \in p} \mathbf{S}_i
\end{equation}
and must be zero for all the ground state configurations. This can be satisfied if we consider four parallel lines composed of four spins belonging to a plaquette $p$, and impose that for each line $A$ the four spins sum to zero,
\begin{equation}
    \sum_{i=1}^{4}\mathbf{S}_{A_i} = 0. \label{eq: degeneracy line sum}
\end{equation}
\begin{figure}[htb]
    \centering
    \includegraphics[width = 1 \linewidth]{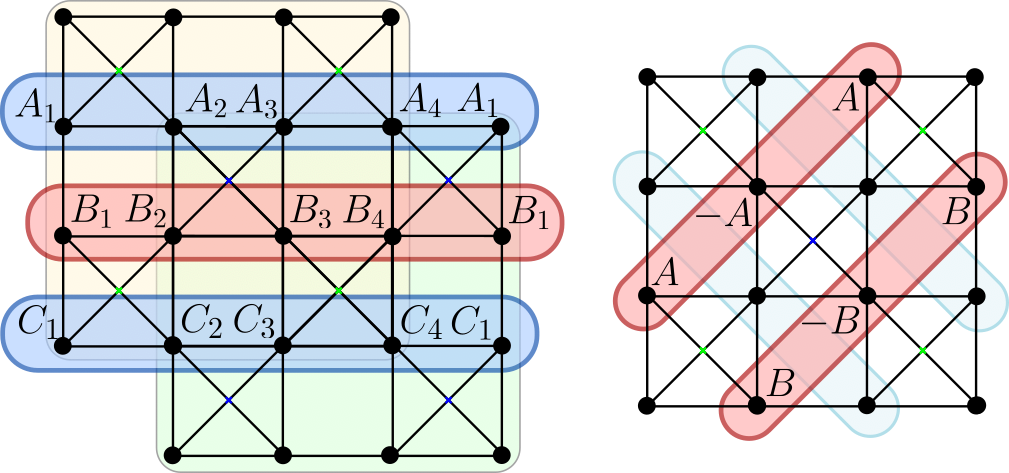}
    \caption{Scheme to illustrate the possible real space configurations associated with the degeneracy lines found in the extended checkerboard model for $(\gamma,\delta)=(1,1)$ (left) and $(0.5, 0)$ (right).}
    \label{fig: Degeneracy lines}
\end{figure}

In this situation, we observe that there are no correlations between different lines, introducing an additional degree of freedom. However, due to the overlap of the extended plaquettes, we also notice that the lines must extend over the entire lattice, and be composed of only four distinct spins along their length, as illustrated on the left side of Fig.~\ref{fig: Degeneracy lines}. In this situation, the structure factor component associated with these configurations will be maximal along the lines in this direction and zero along all other directions, meaning we can roughly write it as 
\begin{equation}
    S(\mathbf{r})_{dl} \cong \delta(y) f_{4a}(x),
\end{equation}
with $f_{4a}(x)$ a function having a four sites periodicity. Note that the same correlation function can be obtained considering vertical lines, simply exchanging $x \leftrightarrow y$. The Fourier transform of this structure factor is expected to have the form of Eq.~(\ref{eq : lines S(q)}), with lines located at positions $c_n = n \pi/2a$ with $n$ a non zero integer. Note that the lines at positions $\pm \pi/a$ do not appear in  Fig.~ \ref{fig: ckb pinch lines}. This is due to the fact that the $\pi$ modes correspond to form neighboring spin doublets $\mathbf{S}_{A_i} = - \mathbf{S}_{A_{i+1}}$, for which there is no constraint of repetition along the line. This $\pi$ mode is thus uncorrelated and does not produce any degeneracy line.

For the case, $\delta = - \gamma = 1$ the idea is the same except that this time the spins from one line must be identified two by two as $\mathbf{S}_{A_1} = \mathbf{S}_{A_3}$ and $\mathbf{S}_{A_2} = \mathbf{S}_{A_4}$. There are again no correlations along the direction perpendicular to the line. Along the line, the spin components taken along the bisector between the two spins result in a zero mode, while the spin components taken in the two spin planes but orthogonal to the first spin axis produce a $\pi$ mode. 
This results in the formation of lines of abscissa $0$ or $\pi/a$ in momentum space, as observed on Fig.~\ref{fig: ckb pinch lines}.

In the case $\gamma = 0.5$ and $\delta = 0$, the picture is similar, except that this time we draw diagonal lines including three sites in each extended plaquette as depicted at the right of Fig.~\ref{fig: Degeneracy lines}. These lines are composed of alternating spins such that each triplet on a line $A$ satisfies
\begin{equation}
    \mathbf{S}_A - 2 \gamma\,\mathbf{S}_{A} =0.
\end{equation}
This produces diagonal degeneracy lines lying on the BZ boundary since there is a $\sqrt{2}a$ periodicity along the real space diagonal correlated lines.

These lines have a huge impact on the structure factor due to the semi-extensive degeneracy, which also explains the smaller specific heat observed in the MC simulations, as discussed in subsection \ref{subsec:MC-checkerboard}. It is important to note that these particular values of $\gamma$ and $\delta$ are necessary to construct these types of lines, which explains why we only observe rectilinear degeneracy lines for three points in the parameter space.\\

There are other types of degeneracy lines showing up when $\gamma = \delta$, which appear to be quasi-circular, enclosing either points $M$ or $\Gamma$. These lines appear in the LTA context as circular contact lines and as bright lines in the structure factors obtained in the MC simulations. These lines indicate the formation of quasi-isotropic structures of a size considerably larger than the generalized plaquette. 

We will now shift our focus to the second example, which shares numerous similarities with the checkerboard lattice discussed in this section, but also introduces new physics related to the underlying structure of the kagome lattice.

\section{Example 2: the kagome lattice seen as corner-sharing triangles}
\label{sec: Kagome}

\begin{figure}[tbh]
	\centering \includegraphics[width=0.9\linewidth]{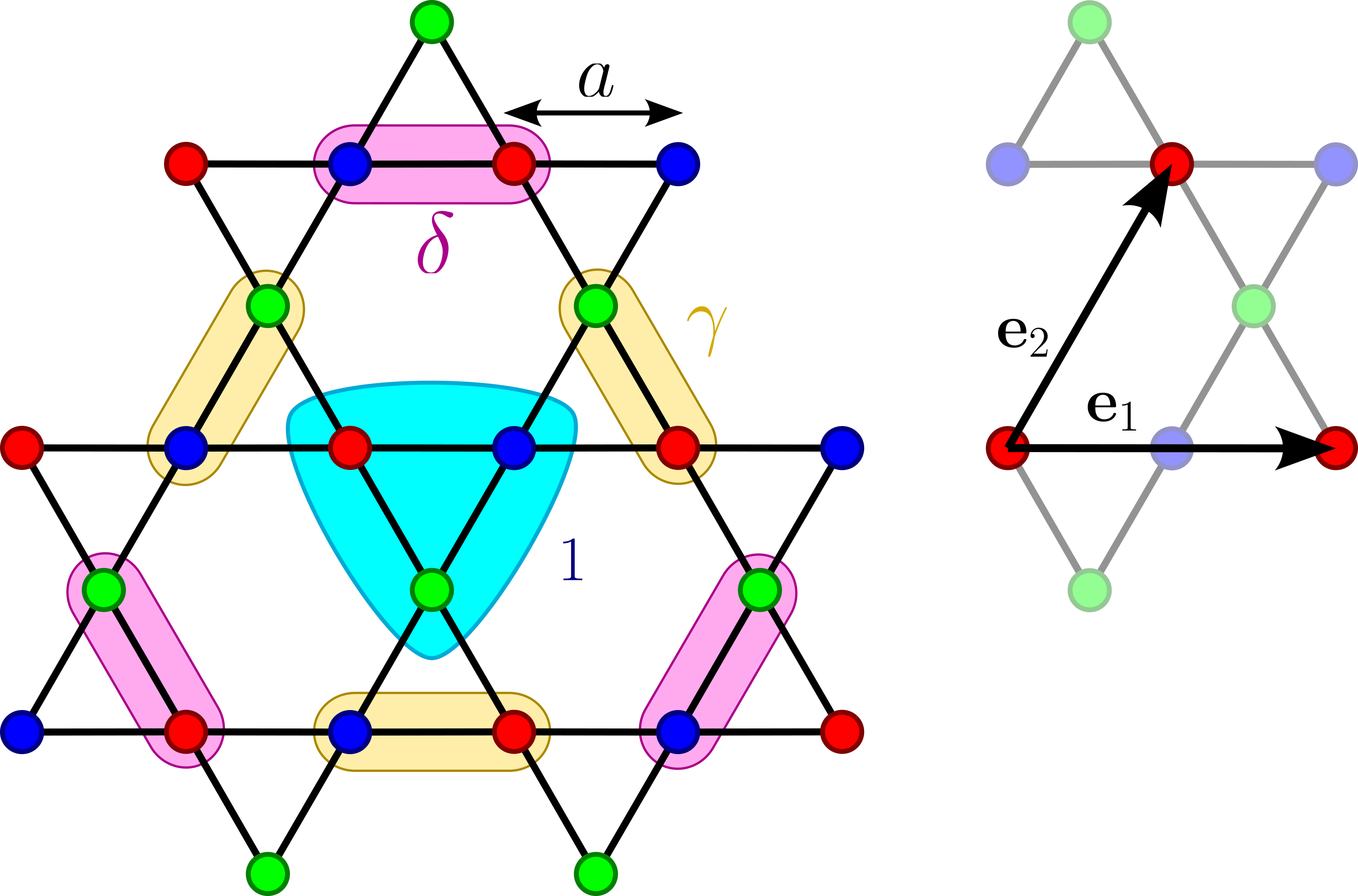}
        \caption{Scheme of extended plaquette for the kagome lattice. First neighbors of standard triangular plaquette come with factor $\gamma$ and second and third neighbors are counted with a factor $\delta$ in the total spin plaquette  definition (Eq.~(\ref{eq: Stot Kagome})). Colors in the sites of the lattice indicate different sublattices. The primitive lattice vectors, denoted by $\mathbf{e}_1$ and $\mathbf{e}_2$, are depicted on the right hand side of the figure.}
	\label{fig: kago2}
\end{figure}

For our second example, we consider an extended triangular plaquette in the kagome lattice, depicted in Fig.~\ref{fig: kago2}, associated with the Hamiltonian in Eq.~(\ref{eq : General Hamiltonian}), where here the total spin in each plaquette $\mathcal{S}_p$ is given by:
\begin{equation}
	\bm{\mathcal{S}}_p = 
		\sum_{i \in p} \mathbf{S}_i + \gamma \sum_{i \in \langle p \rangle} \mathbf{S}_i 
  + \delta \sum_{i \in \langle\langle p \rangle\rangle} \mathbf{S}_i   .
  \label{eq: Stot Kagome}
\end{equation}

 The first sum concerns the spin of the central triangle of Fig.~\ref{fig: kago2} (cyan region), while the second and third sums concern respectively first (yellow region) and second (pink region) nearest neighbor spins. 

\subsection{Luttinger-Tisza approximation}

%
\begin{figure}[th]
	\centering 
 \includegraphics[width = \linewidth ]{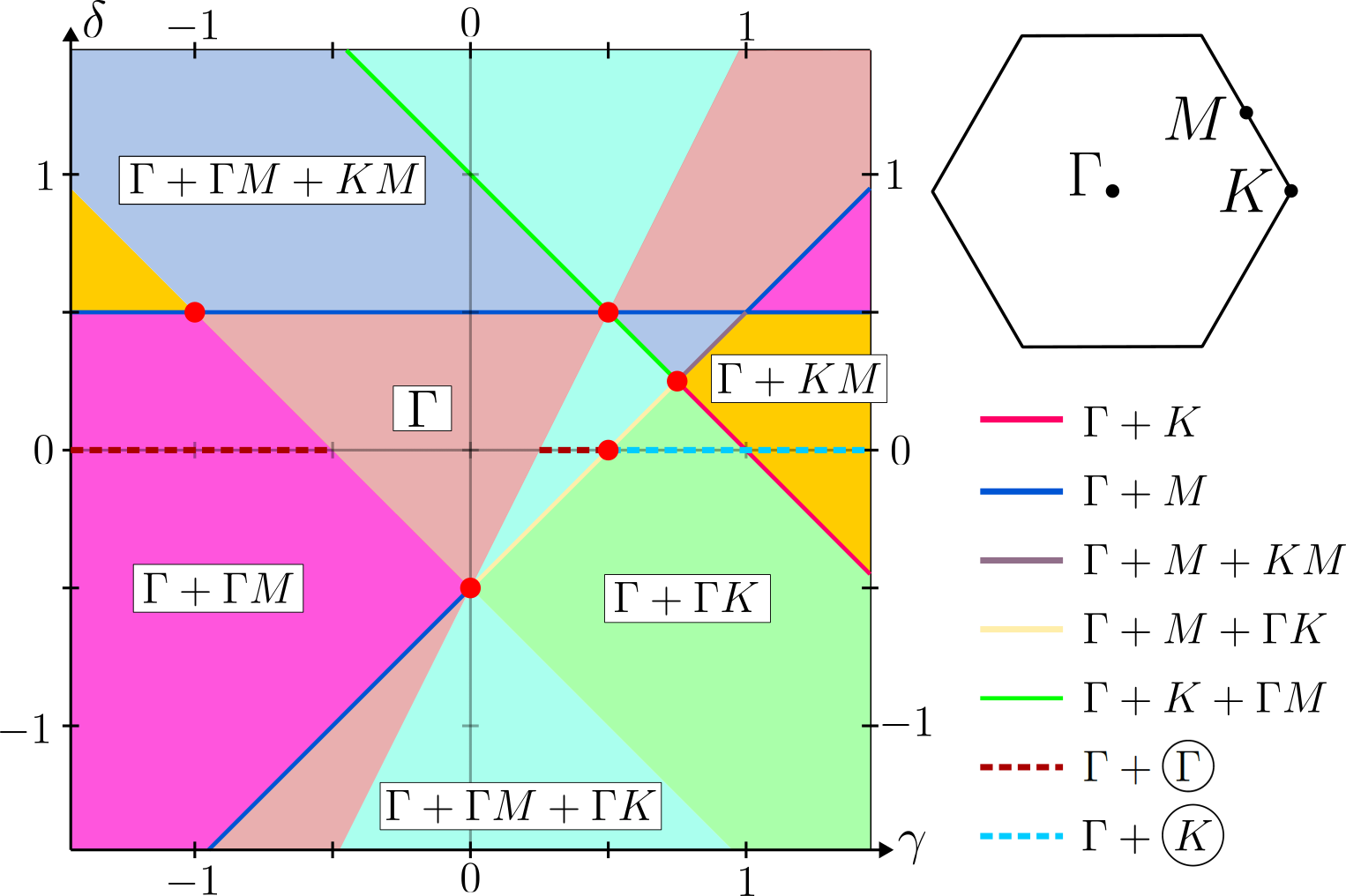}
	\caption{Phase diagram for the Kagome lattice seen as corner sharing extended triangles. This diagram holds five extended phases containing only pinch points. These phases are separated via critical lines with pinch points located on either special point $M$ or $K$. The line $\delta = 0$ presents special features with the appearance of degeneracy lines encircling either $K$ or $\Gamma$ points. There are five special points showing exotic features, denoted by red dots on the phase diagram. } 
	\label{fig: K2 phase diagram}  
\end{figure}

In the kagome lattice, there are three inequivalent sites in the unit cell and lack inversion symmetry with respect to the center of the plaquette. Furthermore, there are two types of plaquettes, corresponding to up- and down-triangles, that are related by central symmetry. As a result, there exist two complex conjugate constraint vectors $\mathbf{L}_\mathbf{q}$ and $\mathbf{L}_\mathbf{q}^*$.

The LTA analysis of the kagome lattice reveals three bands, but only the first band $\varepsilon_1(\mathbf{q})$ is flat (as described in Sec.~\ref{subsec: complex conjugate constraint vectors}). The second band $\varepsilon_2(\mathbf{q}) = (\mathbf{L}^*\cdot \mathbf{L})^2 - (\mathbf{L}\cdot \mathbf{L})^2$ always appears to touch the flat band, similar to the checkerboard case. 
Through an analysis of the location of the points in the Brillouin zone (BZ) where the second band touches the lower flat band $\varepsilon_2(\mathbf{q}^*)=\varepsilon_1(\mathbf{q}^*)$, we can construct the phase diagram illustrated in Fig.~\ref{fig: K2 phase diagram}, here with special high symmetry points at $\Gamma$, $M$, and $K$. The notation used to label phases is similar to the one used for the checkerboard example. The phase diagram is similar in essence to the one of this first example, presenting five extended phases presenting only pinch points, located at point $\Gamma$, and on segments $\Gamma M$, $\Gamma K$, and $KM$. 

As for the checkerboard, to go from one extended phase to another a critical line must be crossed. Some of these critical lines  present pinch points located on the special points $K$ or $M$, which is never the case for the extended phases. Since these pinch points must split into sub-pinch points when crossing the critical line, they correspond to high-order pinch points. They can be classified as for the checkerboard by looking at how many pinch points they split, the contact point dispersion, and their shape in the structure factor. First, when looking at the transition from phase $\Gamma$ to $\Gamma + \Gamma M$ through the line of equation $\delta = 1/2 - \gamma$, the central point $\Gamma$ appears to split into itself plus six sub pinch points, when in the analog situation in the checkerboard it was only splitting into four sub pinch points. Note that the dispersion relation at point $\Gamma$ is quite original since it writes naturally as a sixth-order dispersion divided by the wave vector norm. This still allows a low-temperature description associated with a rank three tensor and thus leads to pinch points with structure as Eq.~(\ref{eq: structure pp rang 3}). The line with equation $\delta = -\frac{1}{2} + 2\gamma$ has a high order pinch point located in $\Gamma$ which splits into itself plus twelve sub pinch points along segments $\Gamma M$ and $\Gamma K$. The associated contact point has a sextic dispersion, and thus enters in the formalism developed at Sec.~\ref{subsec: Low-temperature emergent Gauss tensors}. When looking at the critical line of equation $\delta = 1 - \gamma$, three pinch points along $\Gamma K$ segments appear to collapse in $K$ point to again split into three sub pinch points along $KM$ segments. Curiously, the dispersion at point $K$ is quadratic, corresponding at first glance to a usual pinch point. It does however appear that for this special line of the phase diagram, the second dispersive band also touches the flat band at points $K$ with a quadratic dispersion. In this situation the denominator of the structure factor (\ref{eq: structure factor projective method K2}) obtained using the Projective Analysis is of order four, corresponding to a quartic pinch point. The line $\delta = -\frac{1}{2} + \gamma$ hosts a Lifshitz pinch point located in $M$, splitting into two sub-pinch points along $ M \Gamma$ segments. Finally, the line with equation $\delta = \frac{1}{2}$ is associated with a quartic pinch point located at $M$, splitting itself into four sub pinch points along $M\Gamma$ and $M K$ segments. This information is summarized in Table.~ \ref{Table: K2 hopp}.

\begin{table*}[htb]
\centering 
\begin{tabular}{ccccc}
    \hline
    \hline
    \bfseries Line \hspace{0.1cm} & 
    \bfseries Position & 
    \bfseries \hspace{0.4cm} Multiplicity & 
    \bfseries Dispersion & 
    \bfseries Pinch point  \\
    \hline
    \hline
    $\delta = -\frac{1}{2} - \gamma$ & $\Gamma$ & 1+6 & $\varepsilon_2(\mathbf{q}) = \frac{\alpha_i q_i^6 + \beta_1 q_x^4q_y^2 + \beta_2 q_x^2q_y^4}{\|\mathbf{q}\|^2}$ &
    \adjustimage{height=1.99cm,valign=m}{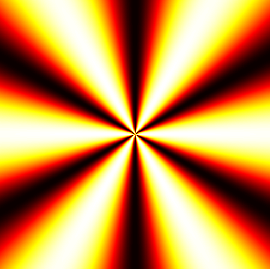} \\
    
    $\delta = -\frac{1}{2} + 2\gamma$ & $\Gamma$ & 1+12 & $\varepsilon_2(\mathbf{q}) = \frac{(1-2\gamma)^2}{8}\|\mathbf{q}\|^6$  & \adjustimage{height=2.0cm,valign=m}{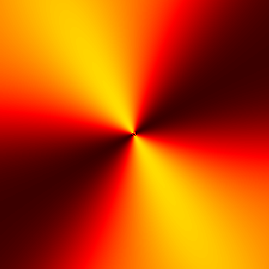} \\
    
    $\delta = 1 - \gamma$ & $K$ & 3 & $\varepsilon_2(\mathbf{q}) \propto \varepsilon_3(\mathbf{q}) \propto \|\mathbf{q}\|^2 $  & \adjustimage{height=1.99cm,valign=m}{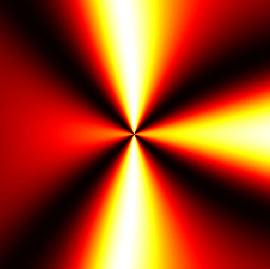} \\
    
    $\delta = -\frac{1}{2} + \gamma$ & $M$ & 2 & $\varepsilon_2(\mathbf{q}) = 2(1-2\gamma)^2 q_x^2  $  & \adjustimage{height=2cm,valign=m}{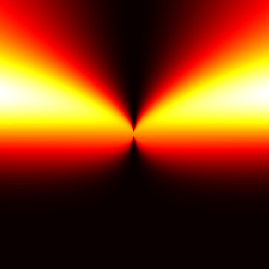} \\
    
    $\delta = \frac{1}{2}$ & $M$ & 4 & $\varepsilon_2(\mathbf{q}) = \alpha_i q_i^4 + \beta\,q_x^2q_y^2 $  & \adjustimage{height=2cm,valign=m}{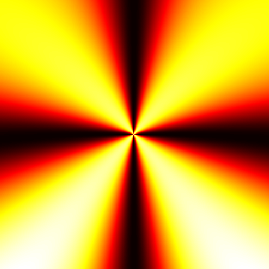} \\
     
\end{tabular}
\caption{Different types of high-order pinch points encountered within the  triangular plaquette kagome model. The dispersion of the band(s) touching at the indicated point is given. The multiplicity indicates in how many points the high-order pinch point  splits when going away from the critical line. The illustration of the structure in the last column is chosen to be representative, but the fine structure of the pinch points changes along a given line. }
\label{Table: K2 hopp}
\end{table*}

A special line in the phase diagram with circular degeneracy lines in the BZ can be observed for $\delta=0$. These degeneracy lines enclose the point $\Gamma$ for $\gamma \in ]-\infty, -1/2[ \cup ]1/4,1/2[$, and surround point $K$ for $\gamma > 1/2$ (except for $\gamma=1$). For $\gamma$ going from $1/4$ to $1/2$, the line encircling the point $\Gamma$ starts growing and goes from circular to hexagonal as it extends in the BZ. When $\gamma$ reaches $ 1/2$, these lines touch the BZ boundaries and appear to form hexagons. When $\gamma$ goes away from $1/2$ these lines take the shape of triangles encircling $K$ points, which shrink up to become points for $\gamma = 1$. For $\gamma > 1$ these triangular lines grow again around the $K$ points, going back to the hexagonal configuration when $\gamma \to \infty$. For $\gamma < 1/2$, the circular lines around $\Gamma$ grow with $|\gamma|$ until they reach the BZ boundary in the limit $\gamma \to -\infty$. The hexagonal lines phase encountered for $\gamma=1/2$ thus corresponds to both limits $\gamma \to \pm \infty$.
Note that this critical line $\delta = 0$ is similar to the $\delta=\gamma$ line observed for the checkerboard lattice. 

There are five special points located at the intersections of special lines which present exotic features. Three of these points show straight degeneracy lines. The point $(\gamma, \delta) = (1/2,0)$ has degeneracy lines forming hexagons along the $MM'$ segments, while at $(-1,1/2)$ and $(0,-1/2)$, the degeneracy lines form six-legged stars along the $\Gamma M$ segments.

The point $(3/4,1/4)$ corresponds to the phase $\Gamma + M + K$ and is, therefore, the only point holding two high-order pinch points: Lifshitz ones in $M$ and quartic ones in $K$. It is similar in essence to the point $\gamma = 0$, $\delta = -1$ observed for the checkerboard. The last special point at $(1/2,1/2)$ corresponds to a situation where the first dispersive band becomes flat and touches the first flat band everywhere. This point is not equivalent to the previous example and will be discussed in Sec.~\ref{subsec: Multicritical point}.

\subsection{Constraint vector function analysis}

\begin{figure*}[htb]
	\centering 
 \includegraphics[width=0.85\textwidth]{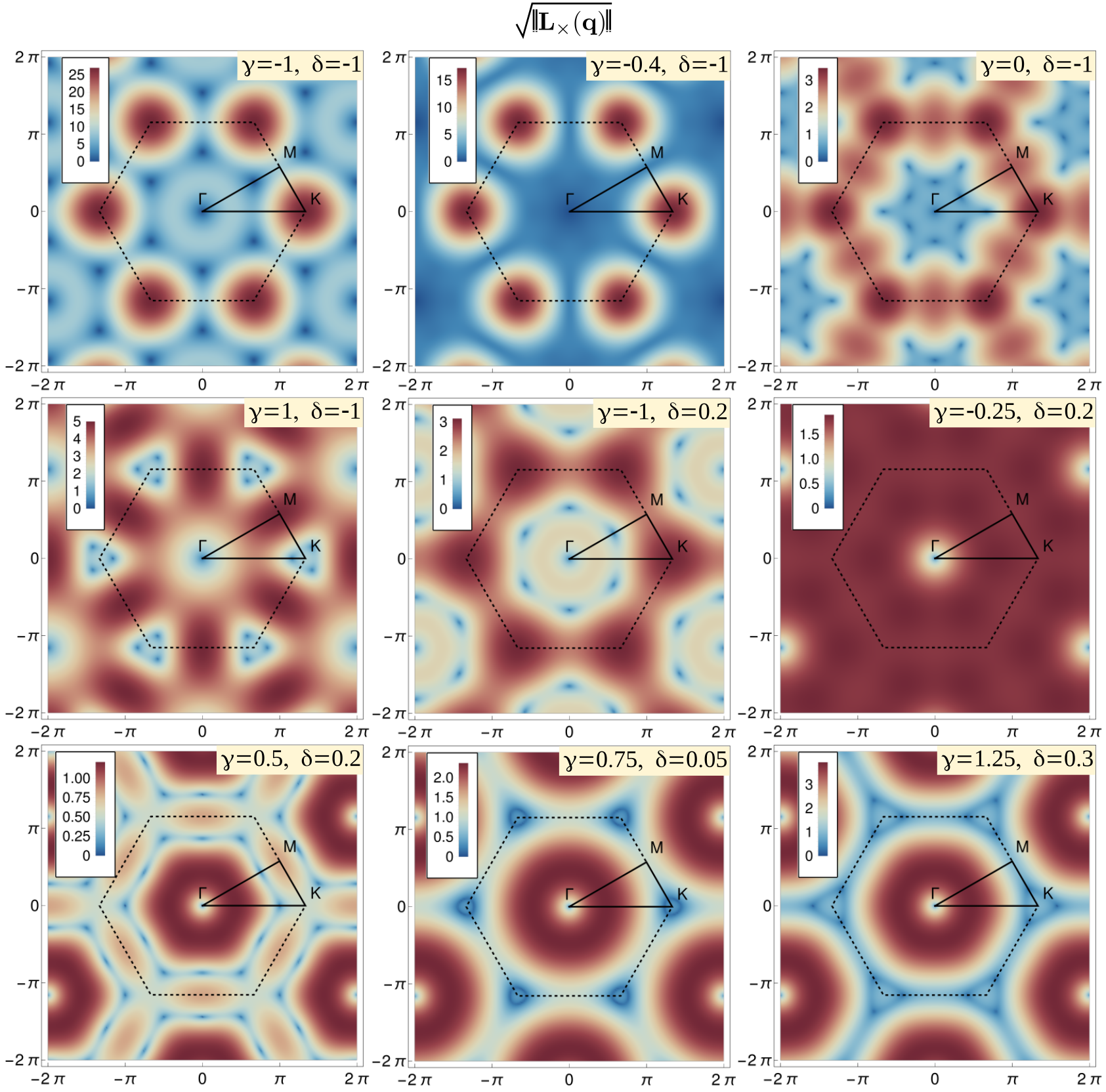}
	\caption{Heat map of $\sqrt{\|\mathbf{L}_\times\|}$ in the Brillouin Zone of the kagome lattice for the triangular plaquette case. We observe that there is a fixed pinch point at the $\Gamma$ point for all cases (where $\|\mathbf{L}_\times\|=0$), while the positions of the other pinch points vary depending on the values of $\gamma$ and $\delta$.} 
	\label{fig:LxLc_Curves}  
\end{figure*}

In this case, the $\mathbf{L_q}$ vector has three complex components. As discussed in subsection \ref{subsec: complex conjugate constraint vectors}, a singularity arises when the real or imaginary part of the constraint vector vanishes, or when they become proportional. Therefore, we focus on the analysis of the real vector field, denoted as $\mathbf{L}_{\times}$ and defined as: $\mathbf{L}_{\times}=i\,\mathbf{L_q}\times\mathbf{L^*_q}$. Figure \ref{fig:LxLc_Curves} displays the plot of $\sqrt{|\mathbf{L}_\times|}$ in the first Brillouin Zone for selected values of $\gamma$ and $\delta$. Our analysis, which uses $\mathbf{L}_\times$ but not $\mathbf{L_q}$, is of course complementary to the LTA. The positions of the zeros of $\mathbf{L}_\times$ can also be observed in the plot. It is worth noting that a pinch point ($\mathbf{L}_\times=0$) is present at the $\Gamma$ point in all cases, while other pinch points emerge at different positions along the $\Gamma-K$, $\Gamma-M$, or $K-M$ paths.

%

\subsection{Monte Carlo Simulation and temperature effects}
\label{subsec: MC K2}

\begin{figure}[htb]
    \centering
    \includegraphics[width=0.9\columnwidth]{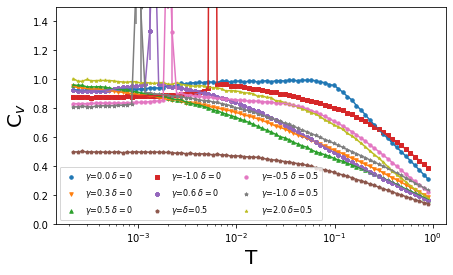}
    \caption{Specific heat per spin as a function of temperature for the extended triangular plaquette kagome model. The temperature is in units of the highest effective coupling. }
    \label{fig:cv_KagTrig}
\end{figure}

Let us now discuss the effect of thermal fluctuations. In the traditional Heisenberg model on the kagome lattice, obtained by setting $\gamma=\delta=0$ in Eq.~(\ref{eq: Stot Kagome}), the coefficients in Eqs.~(\ref{eq:zero-modes}) and (\ref{eq:cv}) are $q=3$ and $b=2$, respectively. However, by including the terms with $\gamma\neq0$ and then $\delta\neq0$, these coefficients jump to $9$ and $6$, and then to $15$ and $10$, respectively, while keeping the ratio $q/b$ constant in all cases. As  $q/b=3/2$, we have $F=0$ in this case, and so the specific heat is expected to be $1$. However, it has been demonstrated that in the simplest scenario with $\gamma=\delta=0$, the specific heat does not behave as expected at lower temperatures. As the temperature decreases, the system goes from a high-temperature paramagnetic phase to a classical algebraic spin-liquid regime, where $C_v\sim1$. Then, as the temperature is further lowered, OBD comes into play. In this case, both analytical and Monte Carlo simulations \cite{OBDkagome,ZhitomirskyPRL2002} have shown that the system selects a submanifold of ground states with soft modes and quartic fluctuations that reduce $C_v$ to $11/12$. In the intermediate spin-liquid regime, the structure factor exhibits the expected pinch points located at the $M$ points of the extended Brillouin zone (EBZ), which are consistent with the LTA, where the two lowest energy bands touch at the $\Gamma$ point. As the temperature continues to decrease, bright peaks associated with OBD selection become visible in reciprocal space \cite{ZhitomirskyPRB2008}.

\begin{figure*}[h!]
    \centering
    \includegraphics[width=1.0\textwidth]{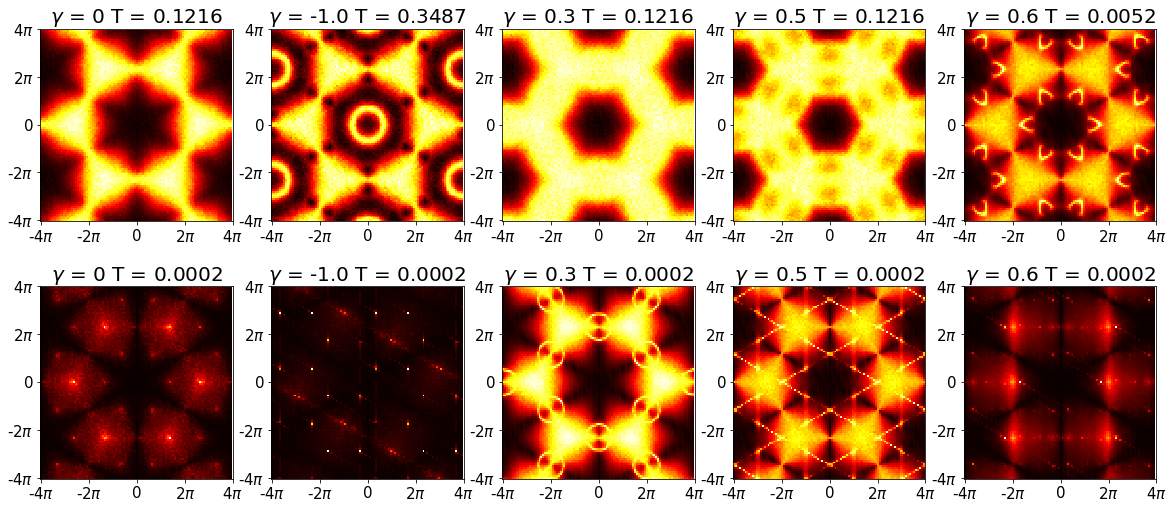}
    \caption{Structure factor for different values of  $\gamma$ in the extended triangular plaquette kagome model obtained with MC simulations, at higher temperatures (top row) and at the lowest simulated temperature (bottom row). }
    \label{fig:Sqs_KagTrig}
\end{figure*}

Now, let's examine the impact of thermal fluctuations in the models that arise by considering $\gamma\neq0$ and $\delta\neq0$, starting with the $\delta=0$ case. Figures~\ref{fig:cv_KagTrig} and \ref{fig:Sqs_KagTrig}  display the $C_v$ vs $T$ plots and $S_\mathbf{q}$ at different $T$, respectively, for various values of $\gamma$, including the well-known $\gamma=0$ case. For all values of $\gamma$,  it can be seen the higher-temperature regime with pinch points in the $M$ points. However, there are some notable differences. Specifically, the $\Gamma + \encircled{\Gamma}$ regime from the LTA analysis differs for $\gamma<-0.5$ compared to $0.25<\gamma<0.5$. In the first case, the $C_v$ at higher temperature seems to tend to a cooperative paramagnet regime, which is consistent with the presence of pinch points in the $S_{\mathbf{q}}$, with an additional bright line encircling the $\Gamma$ point. Then, there is a peak in the $C_v$ indicating a selection, and the $C_v$ is lowered, $C_v<1$, reflected in a change in the $S_{\mathbf{q}}$. In the second case (here we take $\gamma=0.3$), the $C_v$ seems to go monotonically to $1$ as the temperature is lowered, with no additional features, and in the $S_{\mathbf{q}}$ the pinch points are present up to the lowest simulated temperatures ($T=2\times10^{-4}$), and are encircled by bright lines, i. e., there is no state selection, at least in the temperature range we studied.  A similar phenomenon is seen for the special point $\gamma=0.5$, where in this case the pinch points in the $S_q$ coexist with bright ``kagome'' lines that match the LTA predictions. For higher $\gamma$, in the $\Gamma+ \encircled{K}$ case, the $C_v$ and $S_q$ show a similar behavior with temperature as for $\gamma < -0.5$, with the important difference that here the bright lines encircle the $K$ points  of the BZ zone, as predicted by the LTA method. Although the details of the transition from the algebraic spin liquid with degenerate $\encircled{\Gamma}$ and $\encircled{K}$ lines to a lower temperature phase are beyond the scope of this work, this sort of behavior with temperature has been shown in models hosting spiral spin liquids with similar degenerate $\encircled{\Gamma}$ and $\encircled{K}$ lines in the honeycomb and triangular lattices \cite{Okumura2010,Mohylna2022}, where this transition was shown not to be associated with the breaking of any continuous symmetry, consistent with the Mermin-Wagner theorem.

The structure factor obtained with MC simulations can be compared with the structure factor obtained at zero temperature using the Projective Analysis, given by 
\begin{equation}
    S(\mathbf{q}) \propto 1- \sum_{i\neq j} \frac{\|\mathbf{L}\|^2 (L_i^*L_j +c.c) - (Q^* L_iL_j +c.c)}{\|\mathbf{L}\|^4 - |Q|^2}.
    \label{eq: structure factor projective method K2}
\end{equation}

In the special case where $\delta=0$, the difference between the different phases only manifests as the appearance of degeneracy lines, with the only pinch point remaining at the $\Gamma$ point. In such a context, and as it was already noticed in the case of the checkerboard lattice, the projective method fails to reveal the degeneracy lines present in the system. 

\begin{figure*}[ht!]
    \centering
    \includegraphics[width=1.0\textwidth]{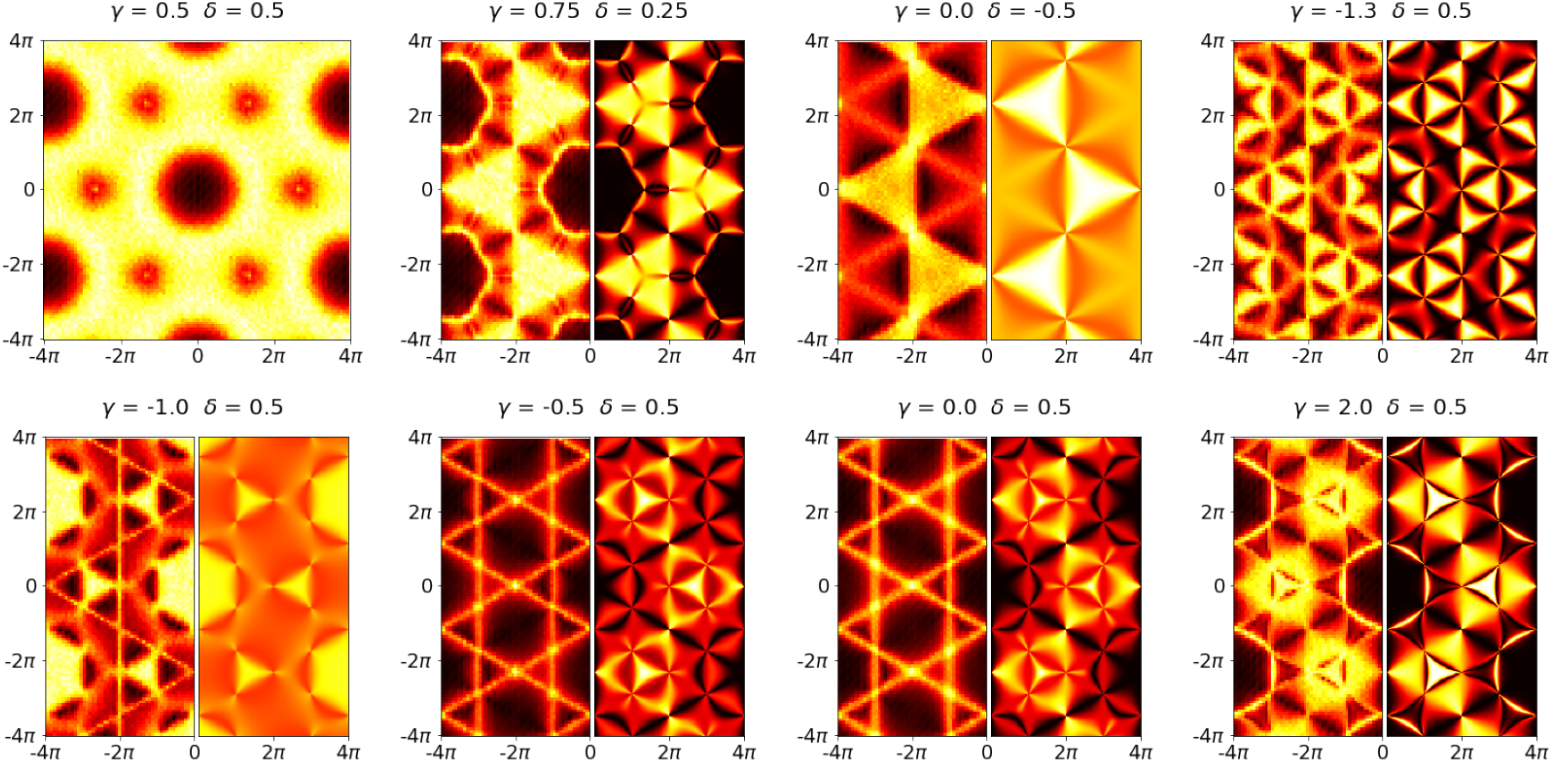}
    \caption{Structure factor for different values of  $\gamma$ and $\delta$ in the extended triangular plaquette kagome model obtained with MC simulations (at $T=0.0052$) and by the Projective Analysis: Except for the case of $\gamma=0.5$ and $\delta=0.5$, in each panel, the left side shows the result from MC simulations while the right side shows the calculation based on PA.}
    \label{fig:Sqs_KagTrig_GD}
\end{figure*}

In the  case of $\delta\neq0$, the LTA shows a variety of phases, including multiple additional pinch-points and higher-order spin liquids. As before, we take some representative cases and perform MC simulations. The resulting structure factors  are shown in Fig. ~\ref{fig:Sqs_KagTrig_GD}.  
 For $(0.75,0.25)$, we see that the algebraic spin liquid holds at the lowest simulated temperature (there is no OBD) and the position of the pinch points matches the LTA prediction and the Projective Analysis, with Lifshitz pinch points in the $M$ points of the BZ and quartic ones in $K$. Something similar is seen at $(-1.3,0.5)$ and $(2,0.5)$, where higher-order pinch points emerge. Then, we present four cases with degenerate lines. In these cases,  at lower temperatures, there is an OBD selection, reflected in the specific heat as a sharp transition that lowers the value of the $C_v$ (see Fig.~\ref{fig:cv_KagTrig}). For $(-1,0.5)$ and $(0,-0.5)$, the degenerate lines match the LTA prediction. However, LTA does not predict degeneracy lines in $(-0.5,0.5)$ and $(0,0.5)$; the reason for this is explained below in Sec. \ref{subsec: degeneracy lines}, which is dedicated to the degeneracy lines.

\subsection{Gauss Law}
\label{sec:Gauss-Law-2}

%
\begin{figure}[htb]
    \centering
    \includegraphics[width=0.6\linewidth]{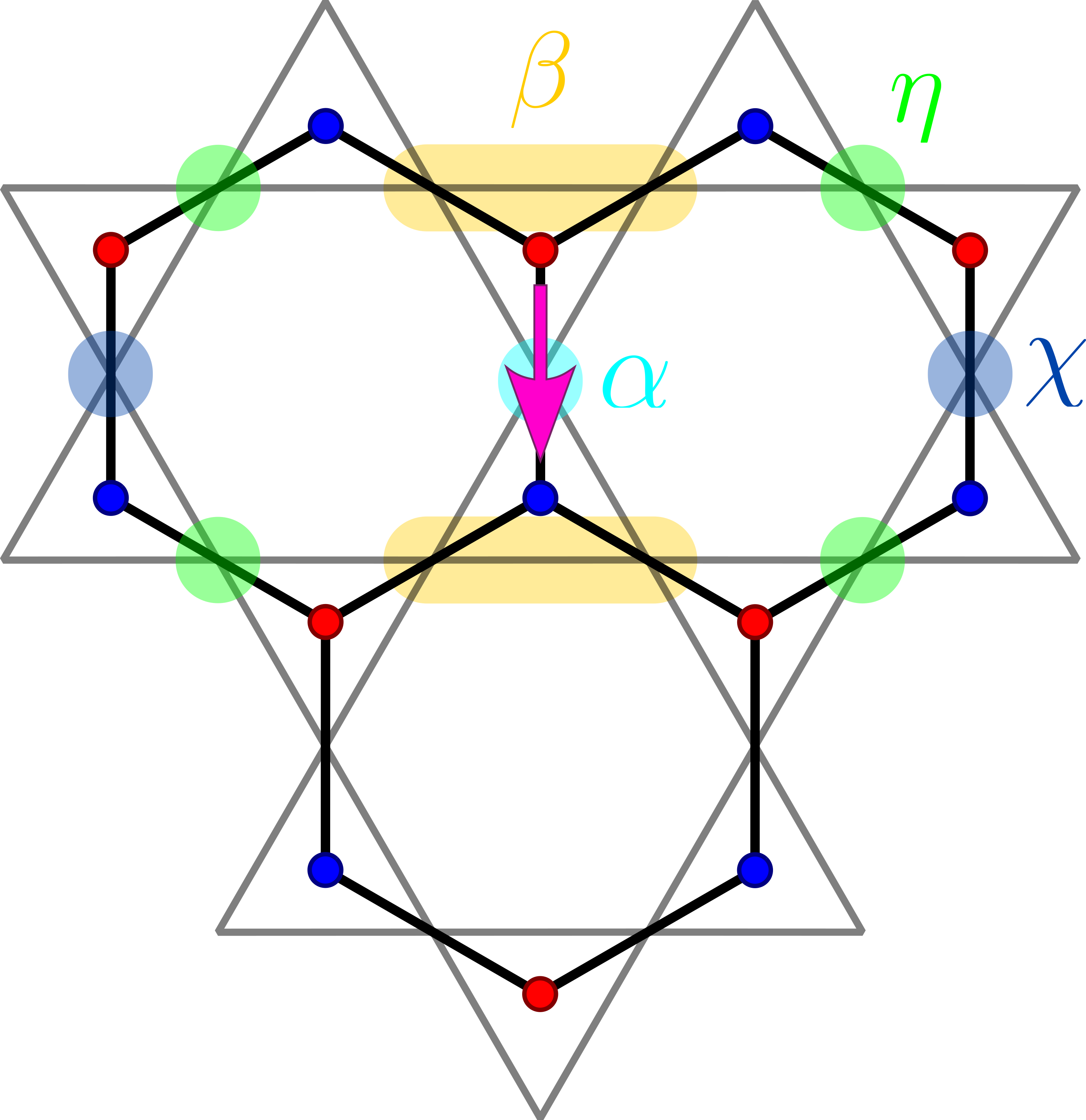}
    \caption{Scheme for constructing the good fluxes for Gauss laws. The bidual lattice of the kagome lattice, that is the honneycomb lattice, is a bipartite lattice. The sites of the two sublattices are represented by red and blue dots. The bonds can then be oriented from red sites towards blue sites, as depicted by the pink arrow. The flux attached to the bond surrounded by a pink arrow is made of the 10 neighboring spins, counted with coefficients $\alpha, \beta, \eta$ and $\chi$ depending on their position relative to the bond.}
    \label{fig: K2 Gauss laws}
\end{figure}
As in the case of the checkerboard lattice, the presence of pinch points for all values of $\gamma$ and $\delta$ brings the question of the existence of a divergence-free polarization tensor that holds true for every combination of these two parameters. To develop this tensor, we follow the same steps as before. We begin with the bipartite honeycomb lattice shown in figure~\ref{fig: K2 Gauss laws}, with its vertices situated at the center of each triangle of the kagome lattice. The bonds of this dual lattice can be oriented and a flux made of neighboring spins can be attached to each of these bonds. The spins which contribute to this flux are the ones belonging to the intersection between neighboring extended plaquettes, and the flux can thus be generally defined using the notations of Fig.~ \ref{fig: K2 Gauss laws} as 
\begin{equation}
    \bm{\Pi}(\mathbf{r}_i) = \alpha \mathbf{S}_i + \beta \sum_{j \in \langle i \rangle} \mathbf{S}_j + \eta \sum_{j \in \langle\langle i \rangle\rangle} \mathbf{S}_j + \chi \sum_{j \in \langle\langle\langle i \rangle\rangle\rangle} \mathbf{S}_j 
    \label{par_kag}
\end{equation}
where the coffecients $\alpha$, $\beta$, $\eta$ and $\chi$ are real.
We want the sum of the incoming fluxes to be equal to the plaquette total spin in Eq.~(\ref{eq: Stot Kagome}), which is zero for any ground state configuration. The coefficients must then be chosen such that the conditions 
\begin{equation}
    \begin{split}
        &\alpha + 2 \beta = 1, \\
        &\beta + \eta = \gamma, \\
        &\eta + \chi = \delta ,
    \end{split}
\end{equation}
are always satisfied. There are $3$ equations to fulfill with $4$ degrees of freedom, meaning that  the relevant fluxes can be built whatever the values of $\gamma$ and $\delta$. The fluxes being well defined, the polarization tensor can then be constructed following Eq.~(\ref{eq: Polarization tensor definition}).

\subsection{Degeneracy lines}
%
\label{subsec: degeneracy lines}

\begin{figure}[htb]
    \centering
    \includegraphics[width = \linewidth]{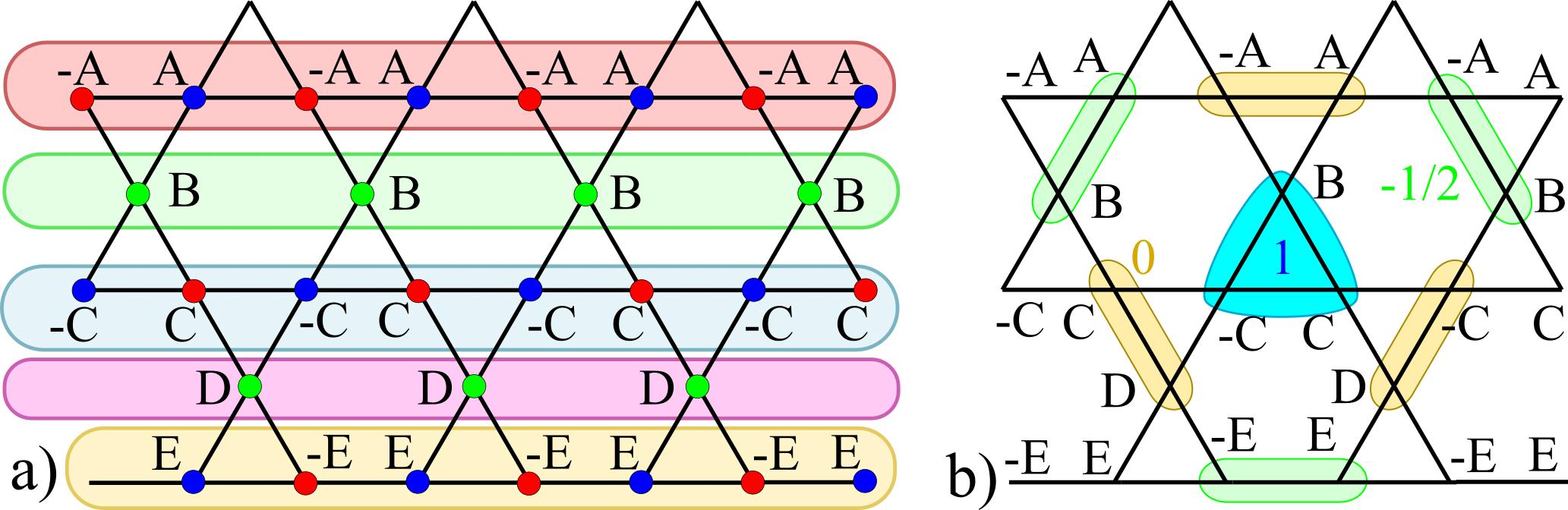}
    \caption{Real space spins configurations responsible for the emergence of vertical bright lines in structure factor for $\gamma = 0$ and $\delta = -1/2$. Figure a) highlights the line structure, showing colored lines of correlated spins along the horizontal direction. Note these lines do not need to have any correlations, as can be seen in Figure b), where it can be checked that this line structure does respect the condition $\bm{\mathcal{S}}_p = 0$ for each extended plaquette $p$. When going from one elementary cell to another the same pattern is reproduced, corresponding to a $0$ spatial frequency.}
    \label{fig: Degeneracy lines K2}
\end{figure}

\begin{figure}[htb]
    \centering
    \includegraphics[width = 0.7 \linewidth]{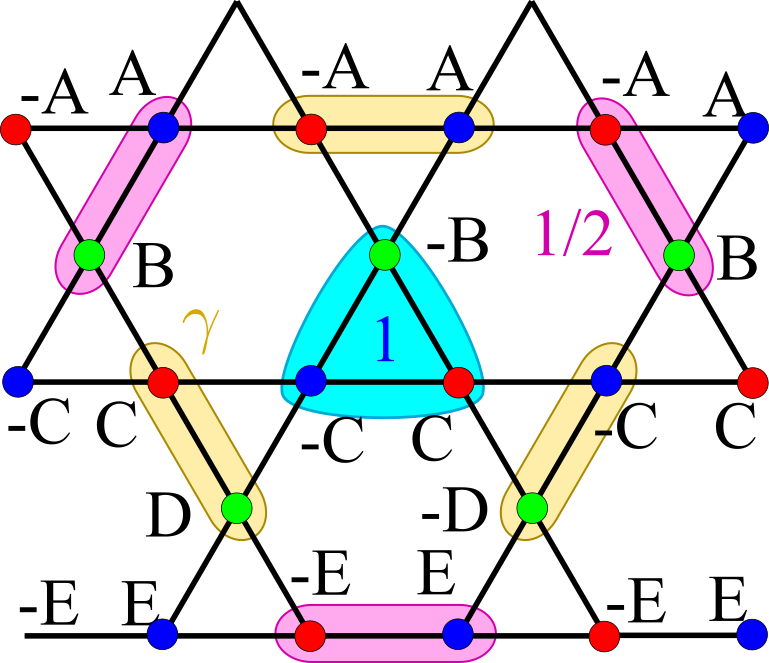}
    \caption{Real space spins configurations responsible for the emergence of vertical bright lines in structure factor for $\delta = 1/2$ with any $\gamma$. Here again, lines of spin can be constructed in such a way that even if these lines are uncorrelated along one direction, they always fulfill the ground state constraint $\bm{\mathcal{S}}_p = 0$ for each extended plaquette $p$. The lines lying on the two first sublattices (red and blue dots) have a $0$ spatial frequency while the one lying on the last sublattice (green dots) has a $4a$ periodicity corresponding to a $\pi/2a$ spatial frequency.}
    \label{fig: Degeneracy lines K2 d05}
\end{figure}

As in the case of the checkerboard lattice, here in the kagome seen as corner-sharing triangles, there are degeneracy lines showing up in the structure factors. Among these lines, some can be observed as contact lines within the LTA, as is the case for $(\gamma, \delta) = (1/2,0), (-1,1/2)$ and $(0,-1/2)$. There are however other lines showing up only in MC simulations, as is the case for $\delta = 1/2$ and any value for $\gamma$. We propose here to present real space configurations corresponding to $\gamma = 0$ and $\delta = -1/2$ on Fig.~ \ref{fig: Degeneracy lines K2} to illustrate the first situation, and the real space construction applying for $\delta = 1/2$ and arbitrary $\gamma$, shown on Fig.~\ref{fig: Degeneracy lines K2 d05}, to discuss the second case. In the first case depicted in Fig.~ \ref{fig: Degeneracy lines K2}, there are lines of correlated spins forming along one direction. These lines are uncorrelated, but share the same spatial frequency along the line direction. This spatial oscillation corresponds here to a zero (or equivalently a $\pi$) mode: when going from one site to the next site belonging to the same sublattice, the spins do not change. This is true for the lines enclosing only sites from the two first sublattices, as for lines only containing sites of the third sublattice. Because these two lines are generally uncorrelated, this means that in the BZ the corresponding bright lines, located at spatial frequency $0$ must correspond to two different modes $(1,-1, 0)^t$ and $(0,0,1)^t$. This means that, on these lines, the subspace of modes with zero energy must have dimension two, implying that these modes belong to different bands, and so the dispersive band must touch the flat one.

The second situation, encountered for $\delta = 1/2$, is different, as depicted in Fig.~\ref{fig: Degeneracy lines K2 d05}. There are again lines of correlated spin extending along one direction and uncorrelated along the orthogonal direction. This time, however, the lines lying on sites of the two first sublattices possess a spatial frequency $\pi/a$, while the lines lying on the third sublattice have a spatial frequency $\pi/2a$. This means each bright line in the BZ relies only on one mode in sublattice space, meaning that the corresponding sublattice subspace attached with zero energy eigenvalue is one dimensional. This means that the degeneracy line can here lie entirely in the flat band, implying no contact line between the dispersive band and the flat one. This line can however appear in MC simulations if an OBD phenomenon selects the submanifold containing real spaces configurations related to one degeneracy line. This is what can be observed in Fig.~\ref{fig:Sqs_KagTrig_GD}, where for $\delta=1/2$ degeneracy lines appear either at frequency $\pi/a$ or $\pi/2a$ depending on the value of $\gamma$ (note that on Fig.~\ref{fig:Sqs_KagTrig_GD} the frequencies are given in $1/2a$ units).

\subsection{Multicritical point}
\label{subsec: Multicritical point}
There exists a special point in the phase diagram, located at $(\gamma, \delta) = (1/2,1/2)$, which has the peculiarity to present two flat bands instead of one in the LTA context. The structure factor observed for these parameters, depicted in Fig.~\ref{fig:Sqs_KagTrig_GD}, is surprising since it does not possess any pinch points. It appears to be identical to the one observed for the next example we discuss, the kagome lattice seen as corner-sharing hexagons. This lattice with a plaquette Hamiltonian (Eq.~(\ref{eq : General Hamiltonian})) was revealed to host  short-range spin liquids\cite{Rehn_Moessner_2017}, and thus presents no pinch points. 

For the special parameters, $(\gamma, \delta) = (1/2,1/2)$ the total spin of an extended triangular plaquette can be expressed as the sum of the total spins of the three hexagons it contains. This means that the ground state manifold of the present model does contain the ground state manifold of the kagome lattice seen as corner-sharing hexagons that we discuss in the next section. It turns out that there is an OBD phenomenon selecting the submanifold of ground states associated with the hexagonal kagome lattice, producing the same structure factor and also a similar specific heat dependence with temperature, see Fig.~\ref{fig:cv_KagTrig}. We checked this by looking at the mean magnetization per hexagons obtained from the MC data: it appears that it is indeed much lower for $(\gamma, \delta) = (1/2,1/2)$ than for other values of the parameters. The fact that the specific heat, in this case, is much lower than the ones obtained for other values of parameters $\gamma$ and $\delta$, being equal to $1/2$ instead of $11/12$, explains the OBD phenomenon: the system chooses to lie in the submanifold presenting the biggest number of soft-modes.

It  Sec.~ \ref{sec:Gauss-Law-2} it has been shown that there is a Gauss law that can be built for every value of $\gamma$ and $ \delta$. On the other hand, it is known, and we discuss this issue in the next section, that the model defined in the hexagonal kagome lattice has short-range correlations without pinch points, which seems in contradiction with the presence of a Gauss law. The resolution of this paradox relies on the fact that for these precise values of the parameters $\gamma$ and $ \delta$, the Gauss law built following the scheme of Sec.~ \ref{sec:Gauss-Law-2} only leads to place closed lines of fluxes wrapped around hexagons of the kagome lattice. More precisely, there is a specific choice for the parameters in Eq.~(\ref{par_kag}) for which the resulting flux field is identically zero. As mentioned before, other choices for these parameters give flux configurations that differ by the presence of flux loops of the shortest length. This means that for any choice in the parameters defined in Eq.~(\ref{par_kag}), the corresponding coarse-grained effective polarization tensor is simply zero everywhere. In such a situation, which corresponds to an effective action like Eq.~(\ref{eq: free energy T0}) with infinite stiffness, the Gauss law, always fulfilled but by a zero field, does not produce dipolar correlations.

\section{Example 3: the kagome lattice seen as corner sharing hexagons}
\label{sec: Corner sharing hexagons}

\begin{figure}[h!]

\centering \includegraphics[width=\linewidth]{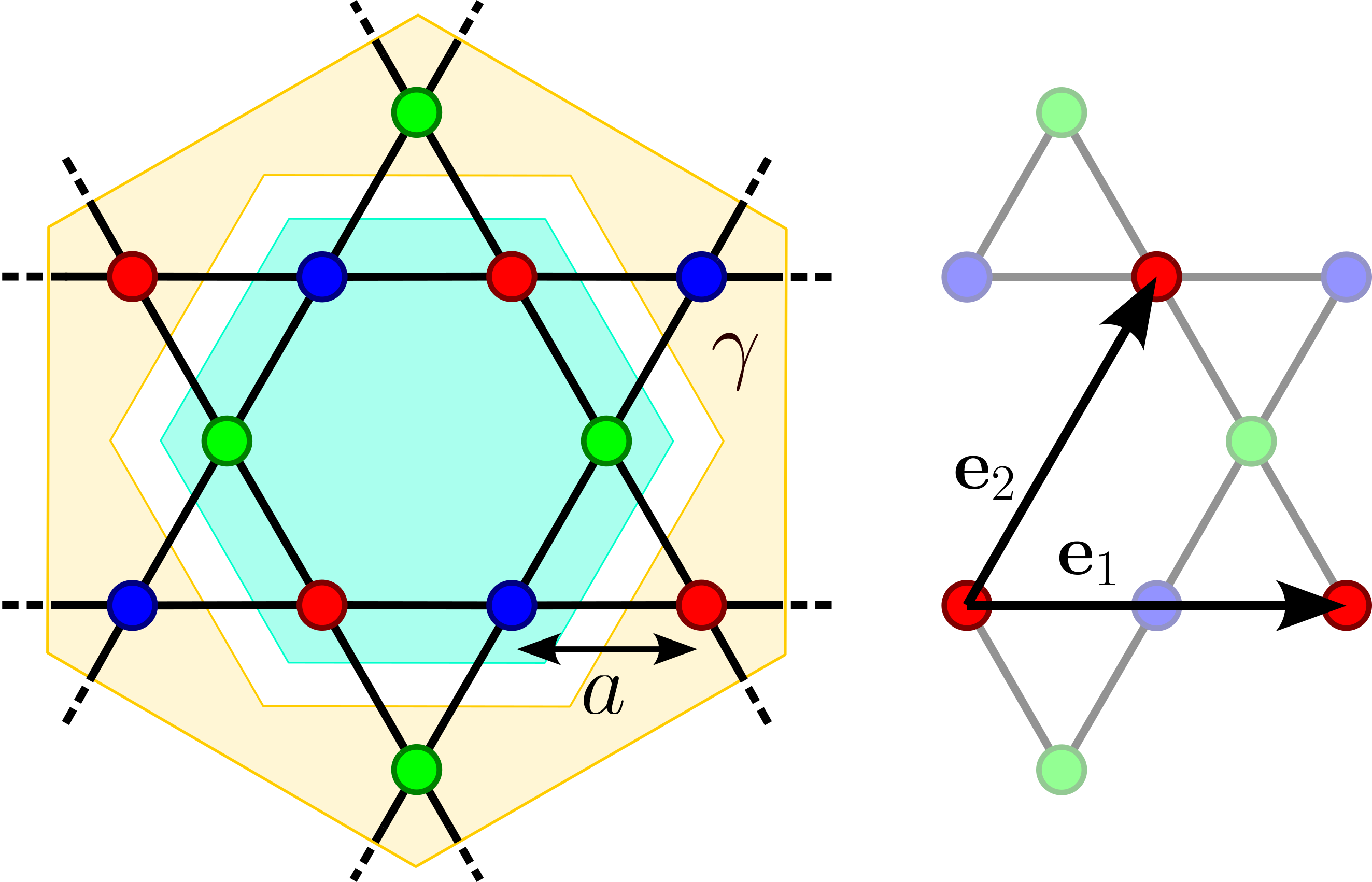}
\caption{Extended model for the corner sharing hexagon plaquette in the kagome lattice. Colors in the sites of the lattice indicate different sublattices. The six central spins belonging to the blue hexagon enter with coefficient $1$ in the plaquette total spin $\bm{\mathcal{S}}_p$ definition and the six surrounding spins are taken with a coefficient $\gamma$. The primitive lattice vectors $\mathbf{e}_1$ and $\mathbf{e}_2$ are are chosen as depicted on the right hand side of the figure.}
 \label{fig: Corner sharing hexagons}
\end{figure}

As the final example in our study, we consider the cluster Hamiltonian in Eq.~(\ref{eq : General Hamiltonian}) for the kagome lattice with extended corner sharing hexagons (see Fig.~\ref{fig: Corner sharing hexagons}), with a total spin per plaquette of 
\begin{equation}
	\bm{\mathcal{S}}_p = 
		\sum_{i \in p} \mathbf{S}_i + \gamma \sum_{i \in \langle p \rangle} \mathbf{S}_i 
\end{equation}
 The second sum accounts for the spins surrounding the central hexagon of the figure. The case $\gamma=0$ was previously examined in \cite{Rehn_Moessner_2017} as an instance of a short-range spin liquid. Note that, in this model, there are three sublattices, which are not equivalent, but the plaquette is symmetric with respect to central inversion.

%

\subsection{Luttinger-Tisza approximation}

\begin{figure}[h]
	\centering \includegraphics[width=0.75\linewidth]{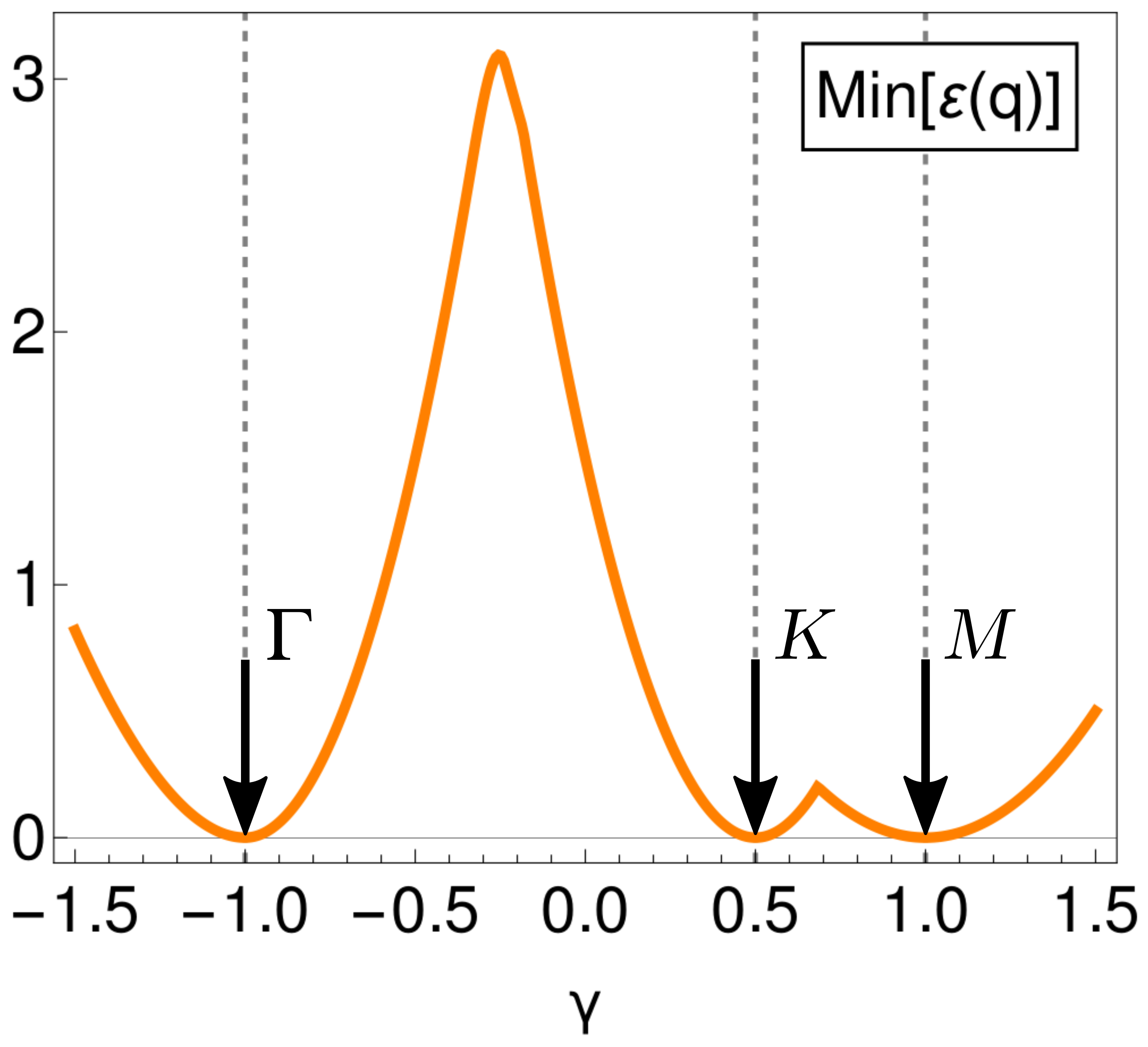}
        \caption{Luttinger-Tisza analysis for the extended kagome lattice as corner sharing hexagons: gap between the third band and the two flat bands  as a function of parameter $\gamma$. The third band  touches the other two flat bands only at three critical values $\gamma= -1, \; \frac{1}{2}$ and 1. Apart from these three points, the third band is gapped.}
        \label{fig: K2 min band}
\end{figure}

As in the previous example for the extended model of the kagome lattice with triangular plaquettes, there are three inequivalent sublattices in this model, but now the plaquette is symmetric with respect to central inversion (see Fig.~\ref{fig: Corner sharing hexagons}).  
Consequently, the vector $\mathbf{L}_\mathbf{q}$ is now a three-component real-valued vector. This implies in the LTA context that there exist two flat bands surmounted by a single dispersive band with dispersion relation $\varepsilon(\mathbf{q}) = \frac{J}{2}\|\mathbf{L}(\mathbf{q})\|^2$. This third band only touches the two flat bands for the specific values of $\gamma$: $-1$, $\frac{1}{2}$ and $1$, see Fig.~ \ref{fig: K2 min band} giving the dispersive band minimum as a function of $\gamma$. In particular, for $\gamma = -1$ the pinch points are  located at the $\Gamma$ points of the first BZ, while the band's contacts are located at the $K$ points for $\gamma = 0.5$ and at the $M$ points for $\gamma = 1$. These results imply that there are four distinct short-range spin liquid regions separated by three points that, as we show below, correspond to algebraic spin liquids.

%
\subsection{Topological properties of the constraint vector function}
%
\begin{figure*}[htb!]
        \centering \includegraphics[width=0.95\textwidth]{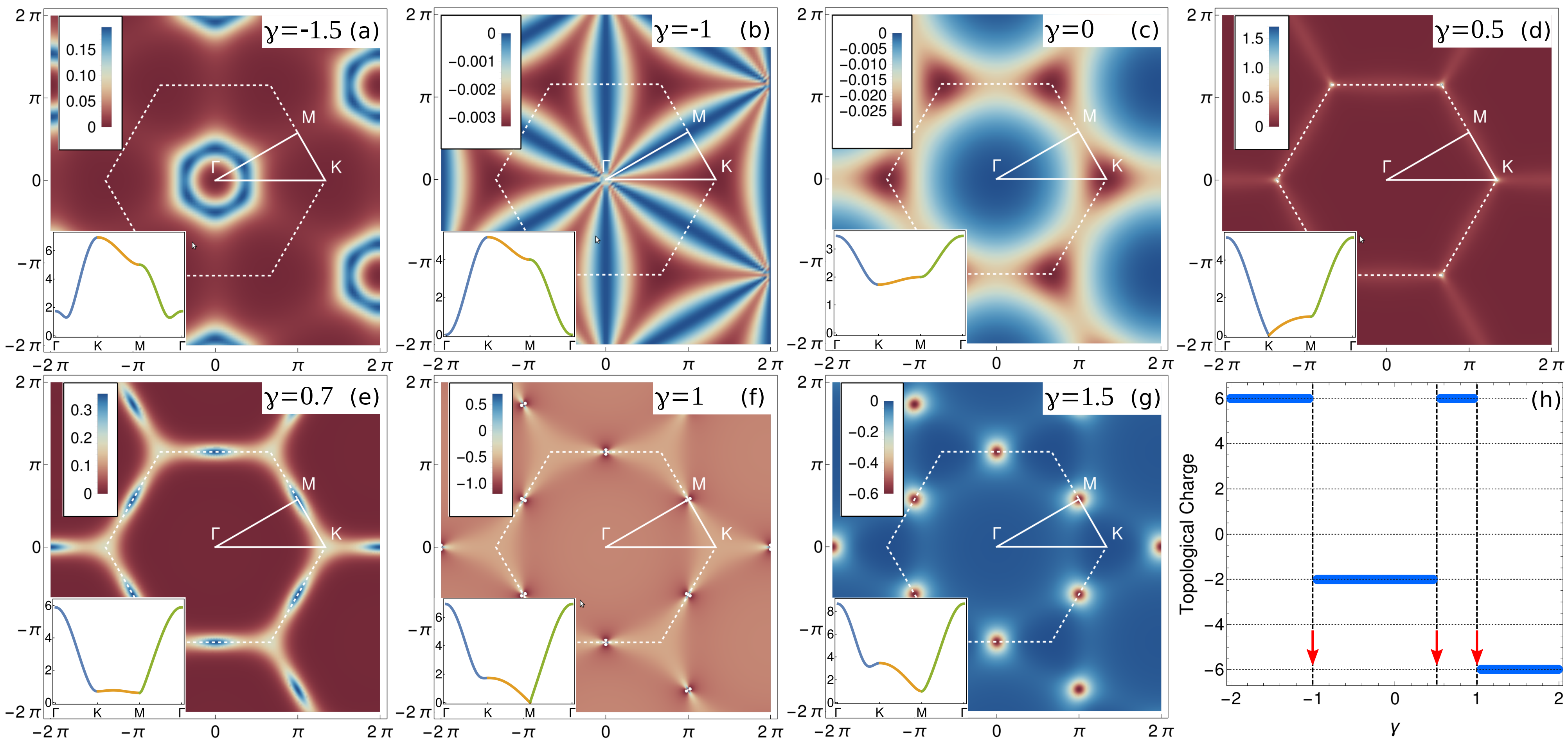}
        \caption{Topological analysis of the constraint vector in the extended hexagonal plaquette model in the kagome lattice. Panels (a-g): density plots of the topological charge for different $\gamma$ parameters. The insets show the absolute value of the constraint vector $\mathbf{L}_\mathbf{q}$ in the $\Gamma-M-K-\Gamma$ line. Panel (h): Total topological charge of the constraint vector integrated over the BZ as a function of $\gamma$. Red arrows indicate the points of the jumps in the topological charge ($\gamma=-1,0.5,1$), which correspond to values where the gap band is closed and pinch points are seen in the structure factor.}
        \label{fig:KagomeHex_Lq}
\end{figure*}
As mentioned before, the constraint vector is now a three-component real-valued vector. It is well known that a three-component vector field defined on the Brillouin zone $\mathbf{L}_\mathbf{q}$ can support topological textures called skyrmions. The associated topological charge of these textures is the total chirality $Q_S$ (or skyrmion number), which in this context is calculated by the integral over the BZ:
\begin{equation}
Q_S=\frac{1}{4\pi}\int_{BZ} {\bf \tilde{L}_q\cdot(\partial_{q_x}\tilde{L}_q\times\partial_{q_y}\tilde{L}_q)}
\end{equation}
where in the definition of $Q_S$ we use the normalized vector ${\bf \tilde{L}_q = L_q/|L_q|}$. The existence of skyrmions (i.e. non zero $Q_S$) does not imply singularities in $\mathbf{L}_\mathbf{q}$; however, the existence of jumps in the total skyrmion number when varying $\gamma$ implies the appearance of singularities necessary to change from one topological sector to another. 

In Fig \ref{fig:KagomeHex_Lq}, panels (a-g) show density plots of the topological charge $Q_S$ for different values of $\gamma$. The insets illustrate the constraint vector norm in the BZ zone, where it can be seen that, as expected from the LTA analysis,  $\mathbf{L}_\mathbf{q}=0$ at different points of the BZ for three values of $\gamma$: $-1$, $0.5$ and $1$.  $Q_S$ as a function of $\gamma$ is shown in panel (h). Each point where the skyrmion number jumps (indicated with red arrows) implies a singularity of $\mathbf{L}_\mathbf{q}$, which in turn indicates a critical point where the band gap is closed and there is an algebraic spin liquid. Therefore, the analysis of the topology of the constraint vector supports the fact that, at the three points $\gamma=-1,0.5,1$, there is an algebraic spin liquids with pinch points separating short-range spin liquid phases.

\subsection{Monte Carlo Simulation and temperature effects}

In the highly frustrated point of the kagome lattice where the Hamiltonian can be expressed as the sum of hexagonal plaquettes, the system is highly degenerate. Each spin is shared by two hexagons, so replacing $q=6$ and $b=2$ in Eq.~(\ref{eq:zero-modes}), it can be seen that there are two zero modes, and the low temperature $C_v$ tends to 0.5. Those values are in principle expected to prevail for $\gamma \neq 0$ as the coefficients $q$ and $b$ become respectively 12 and 4, keeping again their ratio invariant. Previous studies \cite{Rehn_Moessner_2017,AlbaPujol2018} have shown that this particular case is a short-range spin liquid, which is reflected in the structure factor, where no pinch points are seen. As was discussed above, the LTA shows two flat bands and a gapped dispersive band. The extension of the hexagonal plaquette model gives three special points where pinch points may be seen at different points in reciprocal space, for $\gamma=-1,0.5,1$.

\begin{figure}[h!]
    \centering
    \includegraphics[width=0.9\columnwidth]{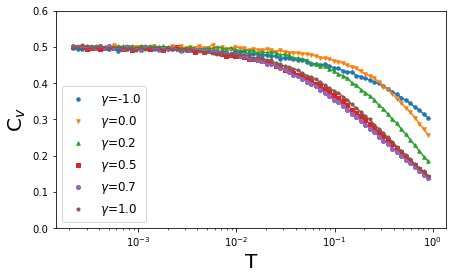}
    \caption{Specific heat per spin as a function of temperature for the extended hexagonal plaquette kagome model. The temperature is in units of the highest effective coupling.}
    \label{fig:cv_KagHex}
\end{figure}

\begin{figure*}[t!]
    \centering
    \includegraphics[width=1.0\textwidth]{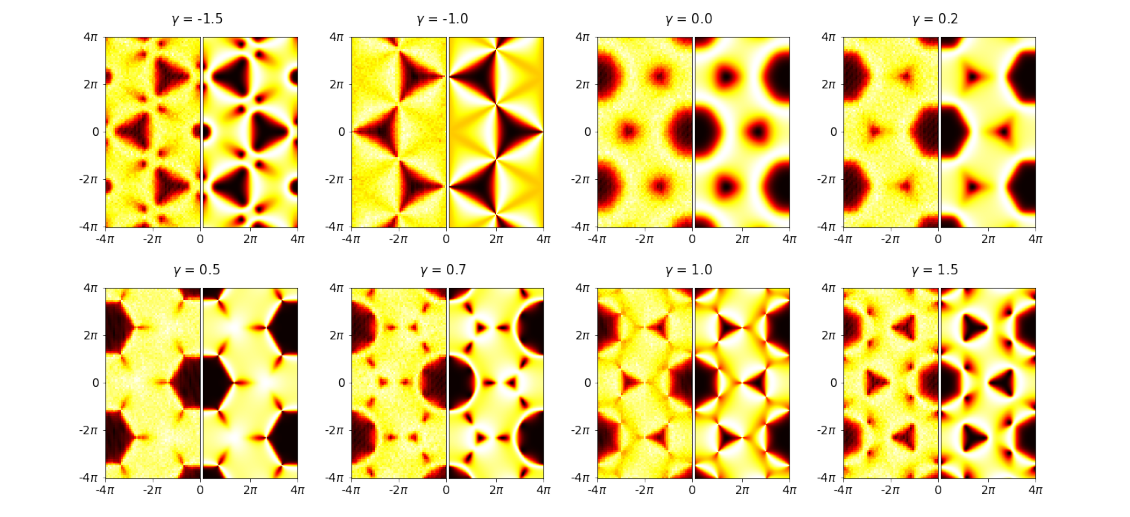}
    \caption{Comparison between the structure factor from MC simulations (at $T=0.0002$) at the lowest simulated temperature (left from each panel) and the Projective Analysis results (right from each panel) for different values of  $\gamma$ in the extended hexagonal plaquette kagome model. }
    \label{fig:Sqs_KagHex}
\end{figure*}

Monte Carlo simulations show that the specific heat at low temperatures goes to $0.5$ for all values of $\gamma$ (see Fig.~\ref{fig:cv_KagHex}). As predicted by the LTA and the constraint vector analysis in the previous subsections, pinch points are clearly seen in the low-temperature structure factors at  $\gamma=-1,0.5,1$, whereas these features are not present for other values of $\gamma$, as shown in Fig.~\ref{fig:Sqs_KagHex}. Comparison with the projective analysis method shows a very good agreement. At $\gamma=-1$, the structure factor is similar to that seen in other models in the kagome lattice, such as chiral spin liquid models \cite{Essafi2016,Rosales2023}, with pinch points in the M points of the EBZ. The position of these features changes in the other two special points, as discussed in the LTA analysis: they are at the K points of the BZ at $\gamma=0.5$ and at the M points of the BZ at $\gamma=1.0$. Therefore, the MC results support the LTA and constraint vector function analysis, evidencing that indeed at low temperatures the extended hexagon plaquette model in the kagome lattice hosts a family of short-range spin liquids separated by three algebraic spin liquids which are distinguishable in reciprocal space.

%

\subsection{Gauss Law}
\label{sec:Gauss-Law-3}

The existence of pinch points for $\gamma = -1,\,0.5$ and $1$ suggests again  looking for an underlying divergent-free tensor for these three special values. We propose here a construction giving a polarization tensor satisfying a Gauss Law for the $\gamma= -1$ and $\gamma=1/2$ cases. Although there might be one also for the case $\gamma = 1$, we did not find an explicit construction in real space and defer this particular case for future studies. 

\subsubsection{The case $\gamma = 1/2$}

The first step consists in showing that this specific case maps to a similar model, studied in Refs.~\onlinecite{Benton_Moessner_2021, Rehn_Moessner_2016}, and defined in the honeycomb lattice with cluster Hamiltonian:
\begin{equation}
    H = \frac{J}{2} \sum _{\hexagon} \left(\sum_{i \in \hexagon}  \mathbf{S}_i\right)^2 = \frac{J}{2} \sum _{\hexagon} \left(\bm{\mathcal{S}}_{\hexagon}\right)^2 .
    \label{H honeycomb}
\end{equation}
\noindent where the plaquettes are bond-sharing hexagons. Consider the $3$ spins of each of the $6$ triangles surrounding the hexagon of Fig.~\ref{fig: Corner sharing hexagons}. They can be grouped to create the effective spins 
\begin{equation}
    \mathbf{S}_{\triangle} = \sum_{i \in \triangle} \mathbf{S}_i,
\end{equation}
located on the sites of a honeycomb lattice. Making the sum of those new effective spins around a hexagonal plaquette corresponds, in the original corner-sharing hexagons lattice, to sum two times the spins in the inner plaquette and one time those in the crown. This precisely gives exactly two times the total spin $\bm{\mathcal{S}}_p$ of the plaquette $p$ for the special case $\gamma = 1/2$, 
\begin{equation}
    \sum_{\triangle \in \hexagon_p} \mathbf{S}_{\triangle} = 2 \bm{\mathcal{S}}_{p}.
\end{equation}
There is thus a mapping relating the model in the honeycomb lattice discussed in Refs.~\onlinecite{Benton_Moessner_2021, Rehn_Moessner_2016}, which has been shown to host an algebraic spin liquid, even if no Gauss Law has been found previously. Note however that in the present case the effective spins $\mathbf{S}_{\triangle}$ are not normalized, but this has no effect on the construction of the Gauss Law. 
\begin{figure}
    \centering
    \includegraphics[width=0.8\linewidth]{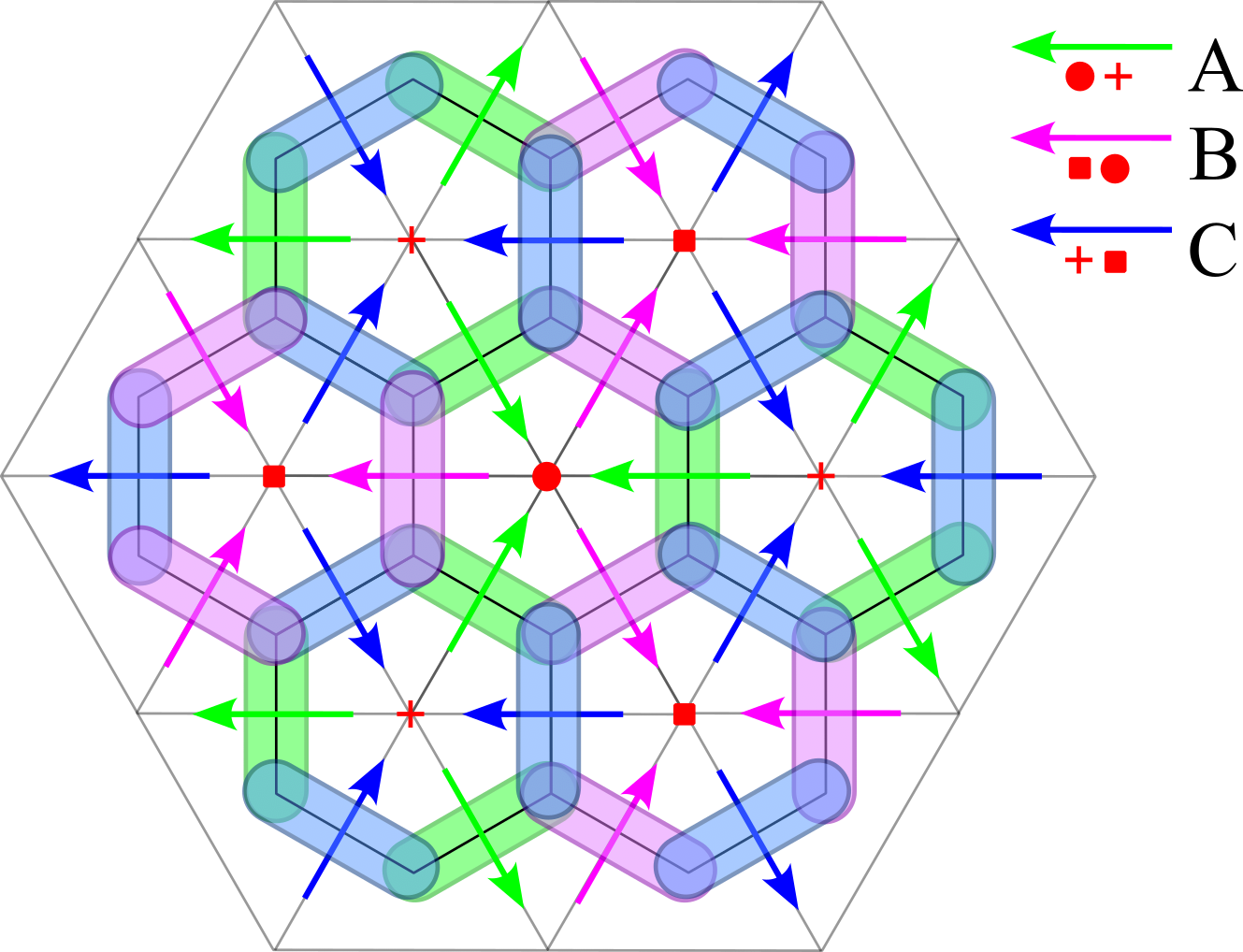}
    \caption{Fluxes construction for the honeycomb model as bond sharing hexagons. The sites of the three sublattices are depicted with the red symbols $\square$, $\circ$ and $+$. The bond are oriented as depicted by arrows. To each of these bonds is attached a flux made of the two neighboring spins.}
    \label{fig: honeycomb}
\end{figure}
We now propose a construction scheme for the polarization tensor in the honeycomb lattice case, a construction that holds for the corner-sharing hexagons lattice presently discussed. Consider the dual lattice of the honeycomb one, which is the triangular lattice. It is a tripartite lattice, meaning $3$ types of sites can be defined in a way such that each type of site doesn't have any neighbors of its own type. These three types of sites are depicted in Fig.~\ref{fig: honeycomb} with little squares, disks, and triangles. In this situation, the lattice bonds can be arbitrarily oriented as depicted in Fig.~\ref{fig: honeycomb}, where a capital letter is associated with each type of oriented bond. We now place on each oriented bond of type $I$ located at position $\mathbf{r}_i$ a flux 
\begin{equation}
    \bm{\Pi}^I(\mathbf{r}_i) = \alpha_I\sum_{j \in \langle i \rangle }\mathbf{S}_j,
\end{equation}
where the coefficient $\alpha_I$ depends on the type of bond considered, and where the sum is made over the two spins sitting aside the bond, see Fig.~\ref{fig: honeycomb}. We now ask that for each type of plaquette $J$, the sum of the incoming and outgoing fluxes  be equal to a number $n_J$ times the vector $\bm{\mathcal{S}}_{\hexagon}^J$, which is $0$ for the ground state configurations. In this way, we ensure that the sum of fluxes entering each vertex is zero. This implies the three relations
\begin{equation}
    \begin{split}
        &\circ \;\; : \hspace{1cm } \alpha_A - \alpha_B = n_{\circ}, \\
        &\square \; : \hspace{1cm } \alpha_B - \alpha_C = n_{\square}, \\
        & + \; : \hspace{1cm} \alpha_C - \alpha_A = n_{+},
     \end{split}
\end{equation}
one for each plaquette type. This system of 3 equations with 6 parameters is not closed. Summing these three equations gives 
\begin{equation}
    \sum_J n_J = 0.
\end{equation}
 This equation is general, with $n$ different parameters $n_J$ for $n$-partite lattices. It does not appear explicitly for bipartite lattices since there are in this case only two different parameters $n_J$ always taken as $1$ and $-1$, $1$ for nodes with incoming links and $-1$ for nodes with outgoing bonds.
For one given choice of $\{n_J\}$ the coefficients $\alpha_I$ can be expressed as
\begin{equation}
    \begin{split}
        \alpha_B &= \alpha_A -n_\circ,\\
        \alpha_C &= \alpha_A + n_+,
    \end{split}
\end{equation}
with $\alpha_A$ remaining here as a free parameter. We can choose for example $n_\circ = n_\square = -1$, $n_+ = 2$ and $\alpha_A = 1$, which imposes $\alpha_B = 2$ and $\alpha_C = 3$. We see that if we consider for example $\circ$ plaquettes and sum the incoming and outgoing fluxes, each spin is entering one time, and outgoing two times, meaning the sum of the fluxes is equal to $-\bm{\mathcal{S}}_{\hexagon}$, and is thus zero as expected.
Once the fluxes have been constructed, the polarization tensor can be defined in a similar way as we did before in Eq.~(\ref{eq: Polarization tensor definition}). Note that the mapping from our model to the bond-sharing hexagon model, and its subsequent Gauss Law, is only possible for the case $\gamma = 1/2$, which plays here the role of an algebraic critical point separating short-range spin liquid phases with no divergence-free polarization tensor. 

\subsubsection{The case $\gamma = -1$}
%
\begin{figure}[htb]
    \centering
    \includegraphics[width = 0.6\linewidth]{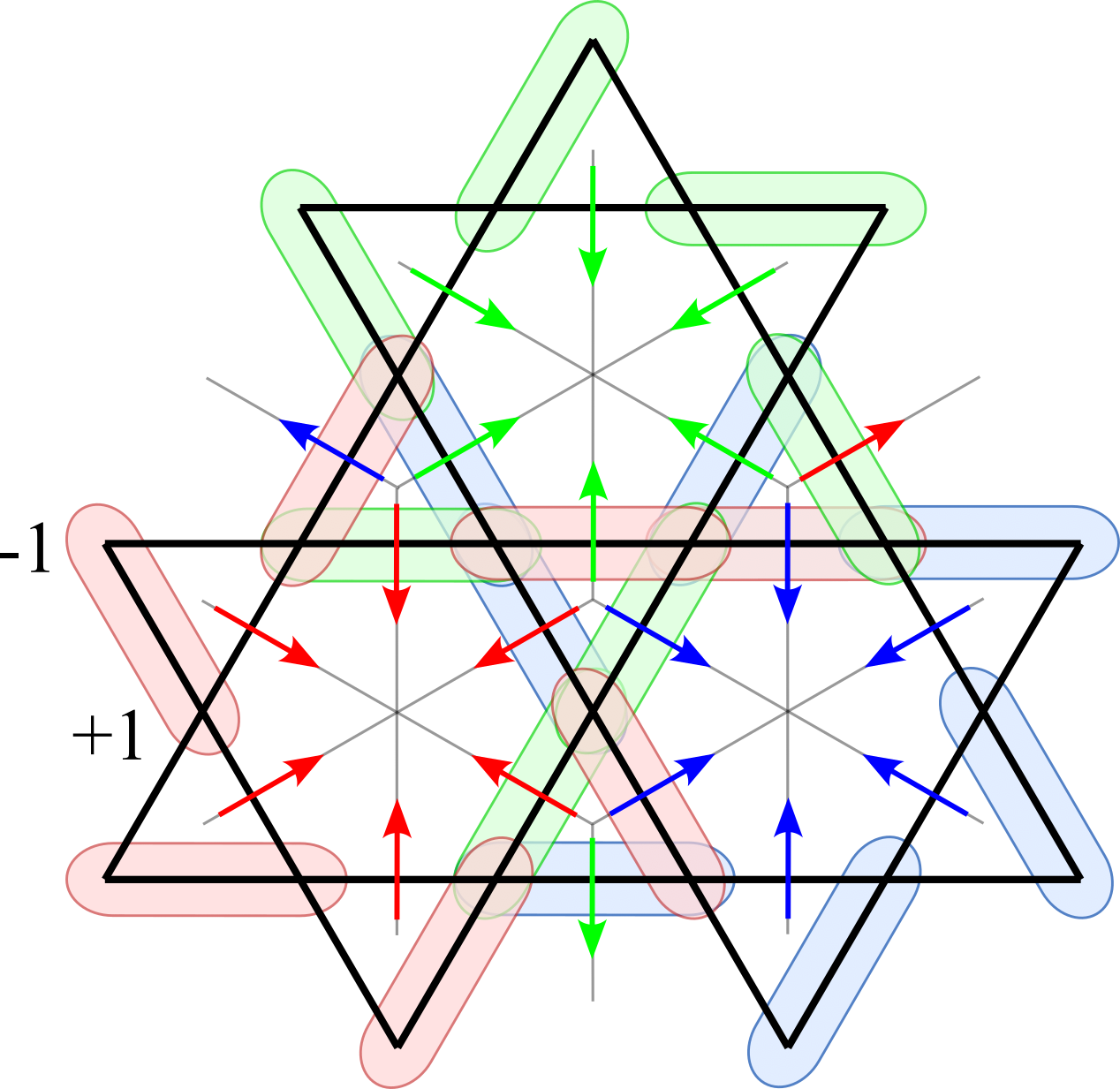}
    \caption{Fluxes structure allowing to build a Gauss Law for corner-sharing hexagons lattice with $\gamma= -1$. Each bond of the dice lattice is oriented as depicted with arrows. A flux made of two spins taken with opposites signs is attached to each bond. }
    \label{fig: Gauss K1 gm1}
\end{figure}
In this case, we can use the kagome dual lattice depicted in Fig.~\ref{fig: Gauss K1 gm1} to place fluxes. Each oriented bond is attached with a flux composed of two spins belonging to the link at the right of the arrow, see Fig.~\ref{fig: Gauss K1 gm1} where the spins attached to a bond are highlighted with the corresponding bond color. The two spins come with signs $+$ or $-$, in such a way that the projection of the dipole on the bond points in the bond direction. In this way we obtain 
\begin{equation}
    \sum _{i \in \hexagon} \bm{\Pi}_i = \bm{\mathcal{S}}_p = 0
\end{equation}
when summing around a hexagonal plaquette. For vertices surrounded by a triangle, the sum of outgoing fluxes gives
\begin{equation}
    \sum_{i \in \triangle } \bm{\Pi}_i = \sum_{i \in \triangle } \left( \mathbf{S}_i - \mathbf{S}_i \right) = 0,
\end{equation}
meaning the sum of the fluxes is zero for each vertex of the dual lattice. The Polarization tensor $\bm{\Pi}_i$ constructed from these fluxes following Eq.~(\ref{eq: Polarization tensor definition}) is thus divergence-free in the ground state manifold, as expected.

\section{Conclusion and perspectives}

In summary, our research provides an extensive investigation of three distinct families of classical spin liquids derived from cluster Hamiltonians, employing a combination of complementary analytical and numerical techniques. On the analytical side, the key ingredient is the definition of the constraint vector $\mathbf{L}_{\mathbf{q}}$, which serves as the building block of the LTA and the Projective Analysis of the structure factors and, in some cases, whose topological properties allow for a classification of the different kind of spin liquid phases. On the numerical side, we used extensive MC simulations which turn out to corroborate the analytical results but also account for the entropic effect at non-zero temperature. 

The first two families of spin liquids that we investigated, defined on the checkerboard and kagome lattices, exhibit algebraic correlations, and a real space Gauss Law is explicitly derived for all parameter values in the Hamiltonian. The analytical and MC analysis also shows that for some critical values of the parameters, higher rank gauge fields emerge. These two models nevertheless differ in the fact that the second one clearly shows OBD phenomena and a subsequent selection within the ground state manifold.  

The final example that we analyzed reveals a distinct qualitative behavior, the system is predominantly in a short-range spin liquid phase for most microscopic parameter values. There are however different short-range phases separated by critical points where the system has algebraic correlations. For two of these points with algebraic behavior, we provided an explicit construction of a divergence-free flux field, and the LTA analysis makes us believe that it should be the case also for the third point. 

In conclusion, our study highlights the effectiveness of complementary analytical and numerical techniques in the investigation of classical spin liquids. By analyzing three distinct families of cluster Hamiltonians, which we can consider as tailored benchmark models, we have shown that we can encompass what we believe is the vast majority of the {\it zoology} that one can encounter in the study of two-dimensional classical spin liquids, such as OBD phenomena, vector and higher rank tensor gauge fields, and its associated multi-arm pinch points, and the co-existence in the phase diagram of short-range and algebraic spin liquid states. 
Our results are also of particular importance from the experimental perspective. Indeed, for example in material candidates for kagome spin liquids such as polymorphs herbertsmithite and kapellasite \cite{PhysRevLett.98.077204, PhysRevLett.100.087202, PhysRevLett.109.037208}, there are competing interactions and it is interesting to see how their corresponding ab-initio Hamiltonians can approach one of the categories of the model studied here. Our results are also of course of first importance for the elaboration of artificial spin liquid materials (see refs. \onlinecite{artificial1, artificial2} for a compelling review of different geometries), where the control on the design of the setup can allow to stick to different families of spin liquids presented here. 
Furthermore, our analysis is not limited to two-dimensional systems, and its extension to three-dimensional models built from extended plaquettes holds considerable promise. 

\section*{ Acknowledgments}

The authors thank the CNRS International Research Project COQSYS for their support. F. A. G. A. and H. D. R. thank the Laboratoire de Physique Th\'eorique, Toulouse, France, for their hospitality during the scientific stay when this project was initiated. F. A. G. A. and H. D. R. acknowledge financial support from CONICET (PIP 2021-11220200101480CO, PIBAA 2872021010
0698CO), Agencia I+D+i (PICT 2020 Serie A 03205), and SECyT-UNLP (I+D X893).

\textbf{Note:} Recently, another independent work regarding the classification of classical spin liquids appeared as a preprint
by Yan et al. \cite{Yan2023}.
\bibliographystyle{apsrev4-1}

%

%

\end{document}